\documentclass[longauth]{aa}
\usepackage{txfonts}
\usepackage{graphicx}
\usepackage{natbib}
\usepackage{subeqnarray}
\usepackage{cases}
\usepackage{ulem}
\usepackage{rotating}
\usepackage{makecell}
\usepackage[colorlinks=true,citecolor=blue]{hyperref}
\begin{document}

\title{Ammonia Observations of Planck Cold Cores}
\author{Dilda Berdikhan \inst{1,2}
\and Jarken Esimbek \inst{1,3,4}
\and Christian Henkel \inst{5,1,6}
\and Jianjun Zhou \inst{1,3,4}
\and Xindi Tang \inst{1,2,3,4}
\and Tie Liu \inst{7}
\and Gang Wu \inst{5}
\and Dalei Li \inst{1,2,3,4}
\and Yuxin He \inst{1,2,3,4}
\and Toktarkhan Komesh \inst{8,9}
\and Kadirya Tursun \inst{1,3,4}
\and Dongdong Zhou \inst{1,3,4}
\and Ernar Imanaly \inst{1,2}
\and Qaynar Jandaolet \inst{1,2}
}

\institute{
Xinjiang Astronomical Observatory, Chinese Academy of Sciences, 830011 Urumqi, P. R. China \\
e-mail: dilda@xao.ac.cn, jarken@xao.ac.cn, tangxindi@xao.ac.cn
\and
University of Chinese Academy of Sciences, 100080 Beijing, P. R. China
\and
Key Laboratory of Radio Astronomy, Chinese Academy of Sciences, Nanjing, JiangSu 210008, P. R. China
\and
Xinjiang Key Laboratory of Radio Astrophysics, Urumqi 830011, PR China  
\and
Max-Planck-Institut f\"ur Radioastronomie, Auf dem H\"ugel 69, 53121 Bonn, Germany 
\and
Astronomy Department, King Abdulaziz University, PO Box 80203, 21589 Jeddah, Saudi Arabia
\and 
Shanghai Astronomical Observatory, Chinese Academy of Sciences, 80
Nandan Road, Shanghai 200030, People’s Republic of China
\and
Energetic Cosmos Laboratory, Nazarbayev University, Astana 010000, Kazakhstan
\and
Institute of Experimental and Theoretical Physics, Al-Farabi Kazakh National University, Almaty 050040, Kazakhstan
}

\abstract
{Single-pointing observations of NH$_{3}$\,(1,1) and (2,2) were conducted towards 672 Planck Early Release Cold Cores (ECCs) using the Nanshan 26-m radio telescope. Out of these sources, a detection rate of 37\% (249 cores) was achieved, with NH$_{3}$\,(1,1) hyperfine structure detected in 187 and NH$_{3}$\,(2,2) emission lines detected in 76 cores. The detection rate of NH$_{3}$ is positively correlated with the continuum emission fluxes at a frequency of 857\,GHz. Among the observed 672 cores, $\sim$22\% have associated stellar and IR objects within the beam size ($\sim$2$\arcmin$). This suggests that most of the cores in our sample may be starless. The kinetic temperatures of the cores range from 8.9 to 20.7\,K, with an average of 12.3\,K, indicating a coupling between gas and dust temperatures. The ammonia column densities range from 3.6\,$\times$\,10$^{14}$ to 6.07\,$\times$\,$10^{15}$\,cm$^{-2}$, with a median value of 2.04\,$\times$\,$10^{15}$\,cm$^{-2}$. The fractional abundances of ammonia range from 0.3 to 9.7\,$\times$\,10$^{-7}$, with an average of 2.7\,$\times$\,10$^{-7}$, which is one order of magnitude larger than that of Massive Star-Forming (MSF) regions and Infrared Dark Clouds (IRDCs). The correlation between thermal and non-thermal velocity dispersion of the NH$_{3}$\,(1,1) inversion transition indicates the dominance of supersonic non-thermal motions in the dense gas traced by NH$_{3}$, and the relationship between these two parameters in Planck cold cores is weaker, with lower values observed for both parameters relative to other samples under our examination. The cumulative distribution shapes of line widths in the Planck cold cores closely resemble those of the dense cores found in regions of Cepheus, and Orion L1630 and L1641, with higher values compared to Ophiuchus. However, the higher line width values in the Planck cold cores, when compared to these dense cores in Ophiuchus, suggest that they might be in more advanced and warmer stages of evolution. Additionally, the line width remains relatively small among the various samples examined. A comparison of NH$_{3}$ line-center velocities with those of $^{13}$CO and C$^{18}$O shows small differences (0.13\,km\,s$^{-1}$ and 0.12\,km\,s$^{-1}$ ), suggesting quiescence on small scales.}

\maketitle
\section{Introduction}
\label{sect:Introduction}
Molecular cores, representing the earliest stages of potential star formation, are crucial for our understanding of the initial conditions of this important process shaping the morphology of galaxies. One approach is to conduct a statistical study of cold dense clumps from unbiased large surveys in the Milky Way. The Planck satellite has carried out the first all-sky survey in the submm-to-mm range, providing a wealth of galactic cold dust cores \citep{2011A&A...536A...1P}. The first release of the all-sky Cold Clump Catalogue of Planck Objects (C3PO) was presented by the \cite{2011A&A...536A..23P}. This comprehensive catalog profiles 10,342 distinctive cold sources that stand out against a warmer background. Within the collection of C3PO clumps, a subset of 915 Early Cold Cores (ECCs) was identified using specific criteria, namely a signal-to-noise ratio (S/N) greater than 15 and color temperatures below 14\,K. The S/N is based on one of the four matched multi-frequency filter (MMF) algorithms \citep{2011A&A...536A...7P, 2012A&A...548A..51M}

This particular ECC sample is included as part of the Planck Early Release Compact Source Catalogue (ERCSC), as presented by the \cite{2011A&A...536A...7P}. The \cite{2016A&A...594A..28P} released the Planck Catalog of Galactic cold clumps (PGCCs) with 13188 sources as the full version of the ECC catalog. These sources, found in various environments, are useful for investigating the initial conditions of star formation, including dynamic processes and the evolution of cores in molecular clouds, making use of their characteristic low temperatures and low levels of activity \citep{2010A&A...518L..93J,2012A&A...541A..12J,2012ApJS..202....4L,2020ApJ...895..119T,2020ApJS..249...33K,2022ApJS..258...17F}.

Plenty of works have been dedicated to the PGCCs using different molecular spectral lines. The Purple Mountain Observatory (PMO) 14m telescope was used to observe $^{12}$CO, $^{13}$CO, and C$^{18}$O transitions in multiple regions. Specifically, 674 cores in Planck cold clumps of the ECCs were observed \citep{2012ApJ...756...76W}, along with 71 cores in Taurus, Perseus, and in the California nebula \citep{2013ApJS..209...37M}, 51 cores in Orion  \citep{2012ApJS..202....4L}, 96 cores in the second quadrant \citep{2016ApJS..224...43Z}, 65 cores in the first Quadrant, and 31 cores in the Anti-Center direction region \citep{2020ApJS..247...29Z}. Additionally, the $J$ = 1-0 transitions of HCO$^{+}$ and HCN \citep{2016ApJ...820...37Y} as well as C$_{2}$H $N$ = 1-0 and N$_{2}$H$^{+}$ $J$ = 1-0, were observed toward the 621 CO selected cores \citep{2019A&A...622A..32L}. Two large projects were also conducted to study PGCCs systematically: the Taeduk Radio Astronomy Observatory (TRAO) mapped the $J$ = 1-0 transitions of $^{12}$CO and $^{13}$CO, while the James Clerk Maxwell Telescope (JCMT) surveyed the 850 $\mu$m continuum emission of more than 1000 PGCCs selected from the TRAO sample in the SCUBA-2 Continuum Observations of Pre-protostellar Evolution (SCOPE) project \citep{2018ApJS..234...28L}. Further studies of several molecular species toward subsets of the SCOPE objects were made using the SMT\,10m, KVN\,21m, Atacama Large Millimeter/submillimeter Array, SMA, NRO\, 45m and Effelsberg 100m telescopes \citep{2016ApJS..222....7L, 2017ApJS..228...12T, 2020ApJ...895..119T, 2020ApJS..249...33K, 2022ApJS..258...17F}. Based on these studies above, it has been discovered that PGCCs demonstrate a state of quiescence and are often believed to represent the earliest stage of star formation.

To characterize the physical properties of the interior of these objects in a large, relatively unbiased sample, observations in dense gas tracers are crucial. Ammonia (NH$_3$) has been established as a reliable dense gas tracer of molecular clouds (e.g., \citealt{1983ARA&A..21..239H,1983A&A...122..164W,1988MNRAS.235..229D}). Specifically, NH$_3$\,(1,1) and (2,2), both belonging to the para-species of ammonia, have been proven to be excellent thermometers at $T_{\rm kin}$\,$<$\,40\,K \citep{1983A&A...122..164W}. Additionally, the critical densities of NH$_3$\,(1,1) and (2,2) are about $10^{3}$\,cm$^{-3}$ \citep{1999ARA&A..37..311E,2015PASP..127..299S}, making them a suitable tracer for dense regions and suffering from minimal freeze-out \citep{2007ARA&A..45..339B}. Therefore, we present a pilot study of NH$_3$\,(1,1), and (2,2) towards ECCs in different environments, complemented by archival CO data, and calculate the characteristics of these cold cores. \citet{2022ApJS..258...17F} conducted a study of PGCCs using the Effelsberg 100m radio antenna to observe the NH$_3$\,(1,1) and (2,2) lines as a supplement to SCOPE targets. The selection of regions for observation was based on two main criteria. Preference was given to regions covered by Herschel observations and to regions characterized by high column density clumps \citep{2018ApJS..234...28L}. In contrast to the SCOPE targets, our sample consists of the whole ECC sample obtained under good observational conditions.

This paper presents a survey of Planck Early Release Cold Cores (ECCs) using ammonia lines, conducted with the Nanshan 26-m telescope. In Sect.\,\ref{sect:Observation}, we provide a detailed description of the sample and observations. In Sect.\,\ref{sect:properties}, we describe the sample properties, and in Sect.\,\ref{sect:results} we present the results of our survey. In Sect.\,\ref{discussion}, a comprehensive discussion of the data is given. Our main findings are summarized in Sect.\,\ref{sect:summary}.

\section{Samples, observations and data reduction}
\label{sect:Observation}
\subsection{The sample}
\label{Source}
\begin{figure*}[h]
\vspace*{0.2mm}
\centering
\includegraphics[width=0.9\textwidth]{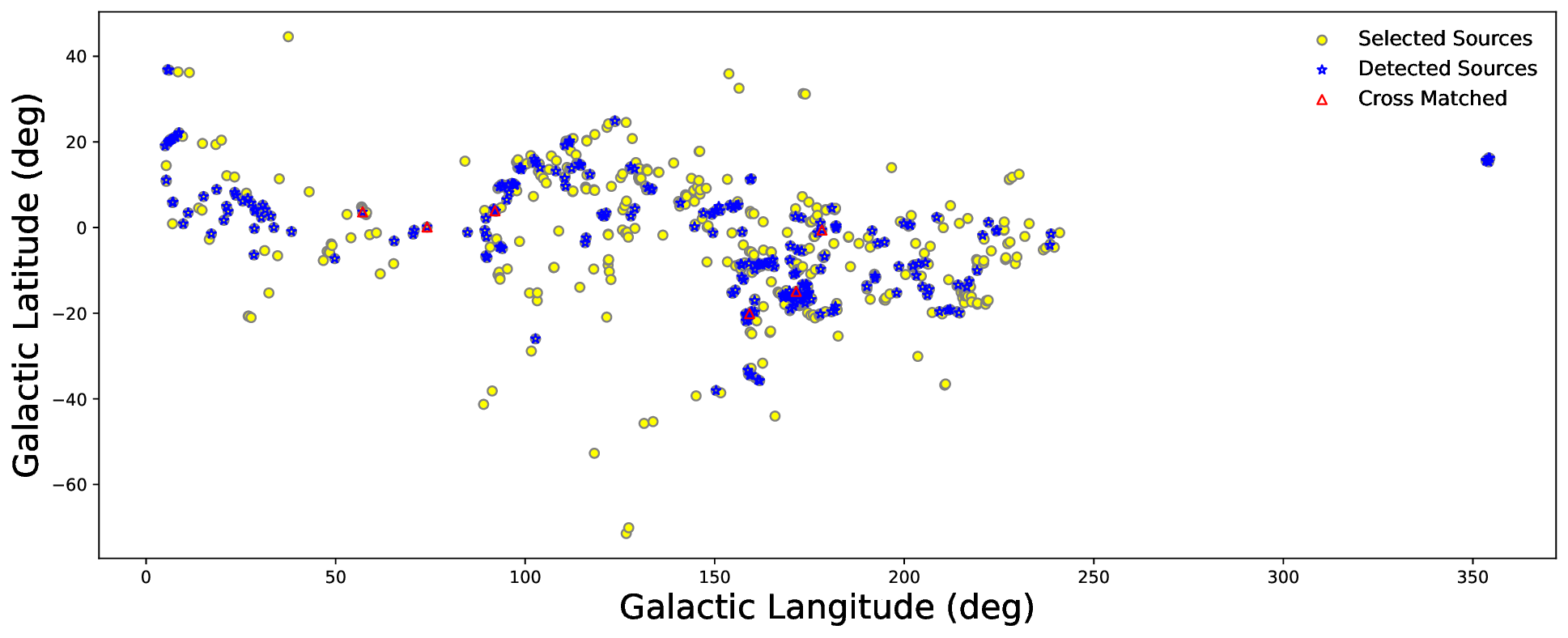}

\includegraphics[width=0.45\textwidth]{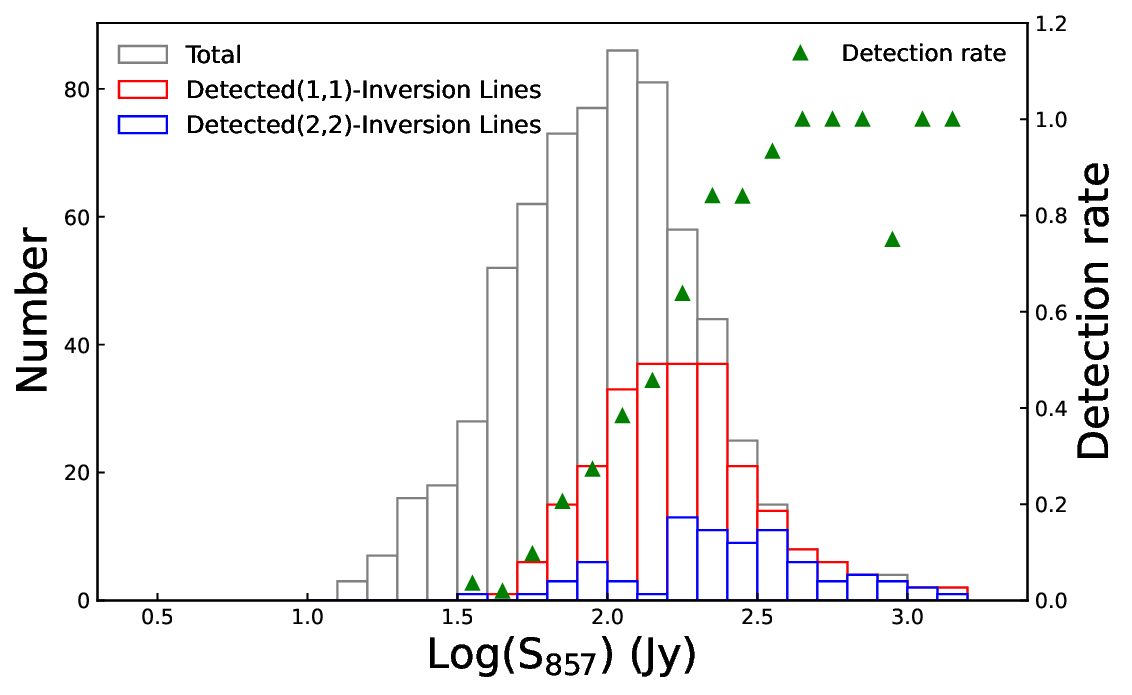}
\includegraphics[width=0.45\textwidth,height=0.27\textwidth]{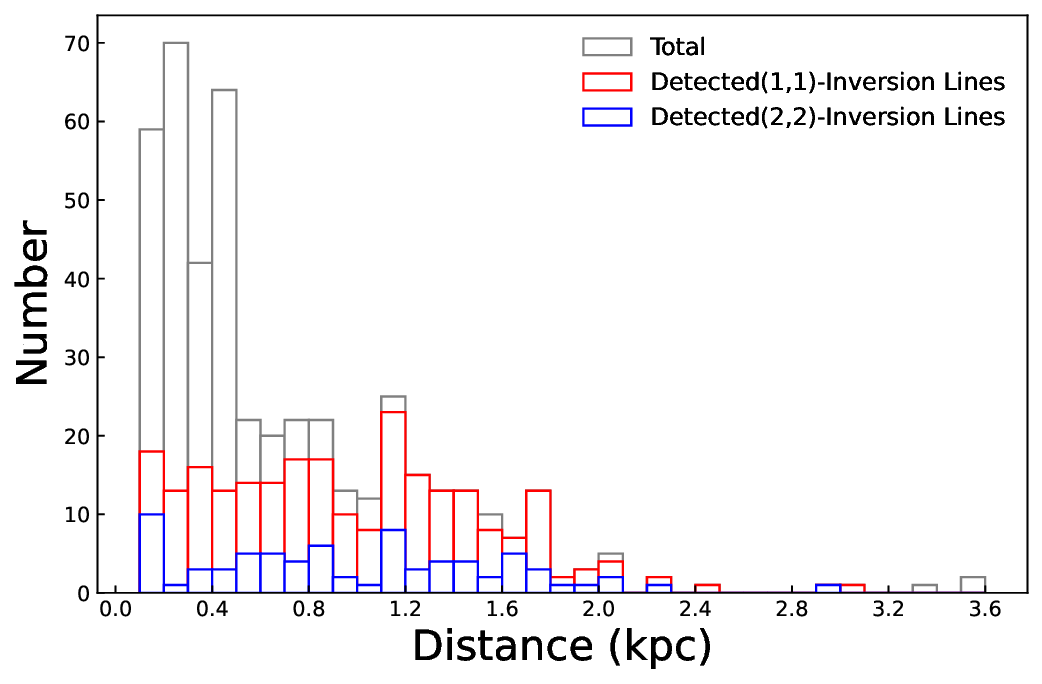}
\caption[]{\textit{Top panel}: The distribution of observed cores in the Galactic plane. The selected cores with and without detections of NH$_{3}$ emission lines are denoted by blue and yellow circles, respectively. The red triangles denote Planck sources not only being part of our sample but having also been previously detected in ammonia by \cite{2022ApJS..258...17F}. {\textit{Lower left}}: Detection rate distribution of NH$_3$ emission lines. The grey histogram represents the whole sample of 672 sources where we searched for NH$_{3}$ lines. The red histogram represents the number of sources that we detected in the NH$_{3}$\,(1,1) line and the blue histogram refers to sources we also detected in the NH$_{3}$\,(2,2) line. The green triangles denote the detection rate for a given 857 $\mu$m flux density. {\textit{lower right}}: Distribution of distances. Here the grey histogram also refers to the entire sample, while the red histogram visualizes those sources detected in NH$_{3}$\,(1,1). The blue histogram refers to sources we also detected in the NH$_{3}$\,(2,2) line.}
\label{position}
\end{figure*}

To ensure optimal observational conditions, we carefully chose cold cores from the 915 Planck Early Release Cold Cores (ECCs) dataset (as previously mentioned in Sec.\,\ref{sect:Introduction}). Specifically, we focused on ECCs with a declination above $-$30 degrees (to observe with a sufficiently high elevation for the Nanshan 26-m telescope). As a result of this selection process, our dataset comprises a total of 672 sources, achieved after the exclusion of sources situated within the Galactic center region, where Galactic longitudes fell within $\left| l \right|$ $\le$ 5 $^\circ$. These exclusions were made due to the complex interstellar medium environment and intense stellar activity prevalent in the central region of the Milky Way \citep{2013A&A...550A.135A, 2016A&A...586A..50G, 2016A&A...595A..94I}. By removing these sources, we aimed to mitigate the potential biases introduced by the unique environment of the Galactic center and obtain a more representative sample of dense cores that better reflect their characteristics in other regions of the galaxy. Subsequently, our observations were conducted with NH$_3$ lines using the Nanshan 26-m telescope. The selected sources are listed in Table\,\ref{table1} of Appendix\,\ref{appendixA} and shown in the \textit{top panel} of Fig.\,\ref{position}. 

Previous investigations have explored extensive sets of individual sources employing ammonia as a probe to examine the dense cores within molecular clouds \citep{1999ApJS..125..161J, 2002ApJ...566..931S, row08, dun11, wie12, 2016ApJ...822...59S}. By conducting a cross-match analysis with these surveys, we identified 27 our sources that overlap with the sample that has been previously detected in ammonia surveys. We also conducted a cross-match analysis with Planck cold cores observed in ammonia \citep{2022ApJS..258...17F}: Only six of our sources have been previously observed in ammonia survey specifically targeting Planck cold cores.

\subsection{NH$_3$ observations}
\label{observations}
We conducted single-point observations of NH$_3$\,(1,1), and (2,2) lines from March 2017 to August 2018 using the Nanshan 26-m telescope. A\,22.0-24.2\,GHz dual polarization channel superheterodyne receiver was employed, with the rest frequency centered at 23.708\,GHz to simultaneously observe NH$_3$\,(1,1) (23.694\,GHz) and (2,2) (23.723\,GHz). 

To convert antenna temperatures $T^\ast_{A}$ into main beam brightness temperatures $T_{\rm MB}$, a beam efficiency of $\sim$0.59 was adopted \citep{2018A&A...616A.111W}, with $T_{\rm MB}$ uncertainty about 14\%. The telescope is equipped with a dual-input Digital Filter Bank (DFB) system with 8192 channels. The typical system temperature is $\sim$50\,K ($T^\ast_{A}$ scale) at 23.708564\,GHz. The FWHM beam of the telescope is 115$\arcsec$ obtained from point-like continuum calibrators, and the bandwidth is 64\,MHz, resulting in a channel spacing of 0.098\,km\,s$^{-1}$. All velocities are with respect to the Local Standard of Rest (LSR). The total integration time for each on-and off-position was 360\,s, with some sources requiring multiple observations to increase the signal-to-noise ratio (S/N). All observations were obtained under good weather conditions and above an elevation of $30\degr$, resulting in a typical RMS noise level of $\sim$20\,mK.

\subsection{CO archival data}
In this study we also utilized the $J$=1-0 transitions of $^{12}$CO, $^{13}$CO, and C$^{18}$O from the PMO \citep{2012ApJS..202....4L, 2013ApJS..209...37M, 2016ApJS..224...43Z, 2018ApJS..236...49Z, 2020ApJS..247...29Z}. The 3$\times$3 beam sideband separation Superconduction Spectroscopic Array Receiver system was used as the front end \citep{2012ITTST...2..593S}. The Half Power Beam Width (HPBW) is 52 arcsec in the 115 GHz band, with a mean beam efficiency of about 50\% and the pointing and tracking accuracies are better than 5$\arcsec$. Fast Fourier Transform Spectrometers (FFTSs) were used with each FFTS providing 16384 channels and a total bandwidth of 1\,GHz. The channel spacing is $\sim$0.16\,km\,s$^{-1}$ for $^{12}$CO, $^{13}$CO and C$^{18}$O. The On-The-Fly (OTF) observing mode was applied, with the antenna continuously scanning a region of \,22$\arcmin$$\times$22$\arcmin$ centered on each clump, while only the central 14$\arcmin$$\times$ 14$\arcmin$ regions were used due to the noisy edges of the OTF maps.

After matching the spatial resolution of previous PMO mapping observations \citep{2012ApJS..202....4L, 2013ApJS..209...37M, 2016ApJS..224...43Z, 2018ApJS..236...49Z, 2020ApJS..247...29Z} with that of the ammonia data (2$\arcmin$), we extracted the spectra of the $J$ = 1 - 0 transitions of $^{12}$CO, $^{13}$CO, and C$^{18}$O at the 2$\arcmin$ $\times$ 2$\arcmin$ peak position of each source in our sample to construct a data set. In the PMO observations, 195 sources match our NH$_3$ detected sub-sample (see Sect.\,\ref{detection_rate}).

\subsection{Data reduction} 
\label{sec-2-4}
The data reduction was performed using the CLASS package of GILDAS\footnote{http://www.iram.fr/IRAMFR/GILDAS/}, and the python plot packages matplotlib \citep{2007CSE.....9...90H}. The NH$_{3}$ data were spectrally smoothed to better compare and analyze these together with the CO data, resulting in a velocity resolution of 0.16\,km\,s$^{-1}$. A feature was considered a genuine detection when the signal-to-noise ratio (S/N) was above 3. To convert hyperfine blended line widths to intrinsic line widths in the NH$_{3}$ inversion spectrum (e.g., \citealp{1998ApJ...504..207B}), we also fitted the averaged spectra using the GILDAS built-in “NH$_{3}$\,(1,1)” fitting method which can fit all 18 hyperfine components simultaneously. From this NH$_{3}$\,(1,1) fit we can obtain integrated intensity, line center velocity, intrinsic line widths of individual hyperfine structure (hfs) components, and optical depth (see Table\,\ref{NH3_line}) of Appendix\ref{appendixA}. 

\begin{figure}[h]
\vspace*{0.2mm}
\centering
\includegraphics[width=0.49\textwidth,height=0.33\textwidth]
{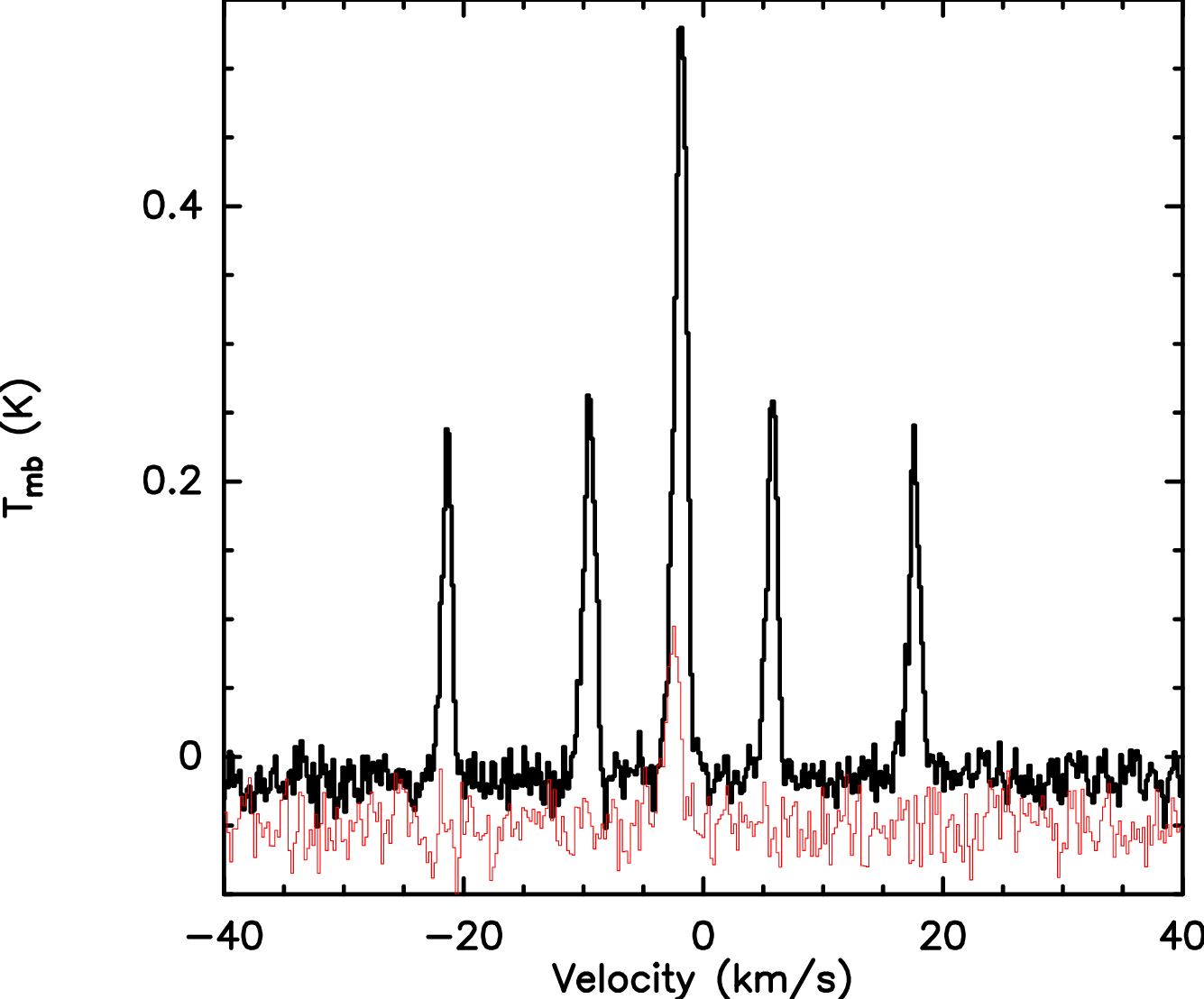}

\caption[]{Example for a typical ammonia spectrum. Shown are NH$_3$ inversion lines from G092.26+03.8. The solid black line represents the observed NH$_{3}$\,(1,1) spectrum and the solid red line indicates the corresponding NH$_3$\,(2,2) profile.}
\label{exempleline}
\end{figure}

Main beam brightness temperatures $T_{\rm MB}$ are obtained from GAUSS fit. Because the hyperfine satellite lines of the NH$_{3}$\,(2,2) transition are mostly weak, NH$_{3}$\,(2,2) optical depths are not determined. A single Gaussian profile was fitted to the main group of NH$_{3}$\,(2,2) hyperfine components. A total of eight NH$_{3}$ cores were fitted with two velocity components. The spectral line of each distinguishable component was fitted with Gaussian and "NH$_3$(1,1)" fittings. These components are denoted by the labels "a" and "b" appended to the respective source names, as presented in Table\,\ref{NH3_line}. We treated these two velocity components as separate entities (i.e., plotted them as two distinct cores). However, most of the cores in our observations required only a single velocity component fit.
Physical parameters of the dense gas such as rotational temperature ($T_{\rm rot}$), kinetic temperature ($T_{\rm kin}$), and NH$_3$ column density ($N_{\rm NH_{3}}$) were derived (see Sects.\,\ref{sect-kinetic_temperature} and\,\ref{NNH3-cal}). Fig.\,\ref{exempleline} provides examples of reduced and calibrated spectra of NH$_3$\,(1,1), and (2,2) inversion lines. Examples for typical ammonia spectrum in different S/N ratios are shown in Fig.\ref{b.1} of Appendix \ref{appendB-spec}

For the $^{12}$CO, $^{13}$CO and C$^{18}$O spectra, peak main-beam brightness temperature, Local Standard of Rest (LSR) velocities, and line widths have been obtained by fitting Gaussian profiles (see Table\,\ref{CO_line} of Appendix\,\ref{appendB}). 
The excitation temperature for CO ($T_{\rm ex (CO)}$), $^{13}$CO opacity ($\tau_{\rm ^{13}CO}$), C$^{18}$O opacity ($\tau_{\rm C^{18}O}$), $^{13}$CO column density ($N_{\rm ^{13}CO}$), C$^{18}$O column density ($N_{\rm C ^{18}O}$), and hydrogen column density ($N\rm _{H_{2}}$) were also calculated (see Table\,\ref{CO_cal} of Appendix\,\ref{appendB}).

\section{Sample properties}
\label{sect:properties}
\subsection{Distances}

The distances of our sources were obtained from the literature \citep{2012ApJ...756...76W,2016A&A...594A..28P}. For sources whose distances were unavailable in the literature, we employed the distance with the highest probability from the parallax-based distance estimator of the Bar and Spiral Structure Legacy Survey \citep{2016ApJ...823...77R}. The histogram of the kinematic distances is presented in the \textit{lower right panel} of Fig.\,\ref{position}. It can be deduced from the distance distribution that the NH$_3$(1,1) detection rate is significantly increased beyond 0.6\,kpc, and the NH$_3$(2,2) detection rate does not exhibit a clear correlation with distance. The distances of all sources range from 0.11 to 4.09\,kpc, with a mean of 0.98\,kpc and a median of 0.94\,kpc. 56\% of the sources have distances within 0.5 and 1.5\,kpc.

\subsection{Kinetic temperature}
With the measured data, the rotational temperature ($T_{\rm rot}$), kinetic temperature ($T_{\rm kin}$), NH$_3$ column density ($N_{\rm NH_{3}}$), thermal velocity dispersion $\sigma_{\rm Therm}$, non-thermal velocity dispersion $\sigma_{\rm NT}$, thermal-to-non-thermal pressure ratio $R_{\rm p}$, thermal sound speed $c_{\rm s}$ and Mach number ($\mathcal{M}$) of cores can be calculated.

\label{sect-kinetic_temperature}
Once the optical depth is determined by the "NH$_3$(1,1)" fitting as described in Sect\,\ref{sec-2-4}, we can calculate the excitation temperature of the NH$_3$\,(1,1) inversion transition through the relation \citep{1983ARA&A..21..239H},
\begin{eqnarray}
\label{Eq1_1}
T_{\mbox{\tiny ex}}=\frac{T_{\rm MB (1,1)} }{\left( 1- \rm exp(-\tau )\right)}\ \ +2.7 \,{\rm K,}
\end{eqnarray}
where $T_{\rm MB}$ and $\tau$ represent the temperature and the optical depth of the (1,1) line derived using the GILDAS built-in `GAUSS' and `NH$_3$\,(1,1)' fitting methods. A histogram of the optical depths of the (1,1) lines for our positions with NH$_3$\,(1,1) signal-to-noise ratios\,$>$\,3$\sigma$ is summarized in Fig.\,\ref{width}b.
 
Since the relative populations of the K = 1 and 2 ladders of NH$_3$ are not directly connected radiatively, they are highly sensitive to collisional processes.
This allows us to use them as a thermometer of the gas kinetic temperature. The method described in \cite{1983ARA&A..21..239H}, has been used to obtain the rotation temperature. 

The rotation temperature is given by the expression

\begin{eqnarray}
\label{Eq1}
T_{\mbox{\tiny rot}}=\frac{-41.5}{\ln\left( \frac{-0.282}{\tau}\ln
\left(1-\frac{T_{\rm MB}(2,2)}{T_{\rm MB}(1,1)}\left( 1- \rm exp(-\tau )\right)
 \right) \right)}\ \  {\rm K,}
\end{eqnarray}
where $T_{\rm MB}$ (2,2) is the main beam brightness temperatures of the (2,2) line derived
using the GILDAS built-in `GAUSS' fitting method.  
 
We estimated the kinetic temperature $T_{\rm kin}$ using the approximation of \citet{2004A&A...416..191T} :
\begin{eqnarray}
\label{Eq2}
T_{\mbox{\tiny kin}}=\frac{T_{\mbox{\tiny rot}}(1,2)}{1-\frac{T_{\mbox{\tiny rot}}(1,2)}
{41.5}\ln\left( 1+1.1 \, {\rm exp}\left(\frac{-16}{T_{\mbox{\tiny rot}}(1,2)}\right)\right)}\ \ {\rm K,}
\end{eqnarray}
where the energy gap between the (1,1) and (2,2) states is $\Delta E_{\rm 12}$= 41.5\,K. This approximation has been derived with Monte Carlo models and provides an accuracy of 5\% in the range between 5 and 20\,K. Most of our sources can be found in this interval.

\subsection{NH$_3$ column density}
\label{NNH3-cal}
Computing the ammonia column density requires the optical depth and line width of the (1, 1) inversion transition along with the rotational temperature, which is obtained from Eqs.\,(\ref{Eq1}). As the optical depth and the rotational temperature depend only on line ratios, the resulting column density is a source-averaged quantity.

Realistically assuming that for our cold sources, the bulk of the ammonia populations resides in the metastable (J = K), (J,K) = (0,0) to (3,3) levels, the total NH$_3$ column densities can be calculated from NH$_3$\,(1,1) following \citep{wie12},
\begin{eqnarray}
\label{Eq3}
N_{\rm {NH_{3}}} \approx N(1,1) \Bigg( \frac{1}{3} \, {\rm exp} \left(\frac{23.1}
{T_{\mbox{\tiny rot}}(1,2)}\right)+1+ \frac{5}{3}\,{\rm exp} \left(-\frac{41.2}{T_{\mbox{\tiny rot}}(1,2)}\right)\\  \nonumber
+\frac{14}{3} \, {\rm exp}\left( -\frac{99.4}{T_{\mbox{\tiny rot}}(1,2)}\right)\Bigg)\ \ {\rm cm^{-2}}
\end{eqnarray}
 and 
\begin{eqnarray}
\label{Eq4}
N(1,1)=6.60\times10^{4}\,\Delta v\,\tau\,\frac{ \,T_{\rm rot}}{\,\nu}\,\ \ {\rm cm^{-2},}
\end{eqnarray}
where $N$(1,1) is the column density of the NH$_3$\,(1,1) transition, the FWHM line width $\Delta v$ is in \,km\,s$^{-1}$, the line frequency $\nu$ is in GHz, and the rotational temperature $T_{\rm rot}$ is in Kelvin. 

\subsection{Velocity dispersions, sound speed, and gas pressure ratio}
\label{sec-dispersion}
The observed line widths provide a measure of the internal motions within each Planck source. 
Here we computed non-thermal velocity dispersion
($\sigma_{\rm NT}$), thermal velocity dispersion ($\sigma_{\rm TH}$), and sound speed ($c_{\rm s}$) following \cite{2014A&A...567A..78L}, \cite{2017A&A...598A..30T, 2018A&A...609A..16T, 2018A&A...611A...6T, 2021A&A...655A..12T}. 
\begin{eqnarray}
\sigma_{\rm NT} = \sqrt{\sigma_{\rm v}^{2}-\sigma_{\rm TH}^{2}}\ \ \,\rm km\,s^{-1},
\end{eqnarray}
where $\sigma_{\rm v}$ = $\Delta v(1,1)$/8ln(2), and $\sigma_{\rm TH}$ = ($k_{\rm B}$$T_{\rm kin}$/17$m_{\rm H}$)$^{1/2}$. $m_{\rm H}$ is the mass of a single hydrogen atom, and $k_{\rm B}$ is the Boltzmann constant. 

The thermal sound speed can be calculated with 
\begin{eqnarray}
c_{\rm s} = \sqrt{\frac{k_{\rm B}T_{\rm kin}}{\mu m_{\rm H}}} \ \ \,\rm km\,s^{-1}.
\end{eqnarray}
where $\mu$ = 2.37 is the mean molecular weight for molecular clouds \citep{2016ApJ...819...66D}.

We also calculated the thermal-to-nonthermal pressure ratio
($R_{\rm P}$=$\sigma_{\rm TH}^{2}$/$\sigma_{\rm NT}^{2}$; \citealt{2003ApJ...586..286L})
and Mach number (given as $\mathcal{M}$=$\sigma_{\rm NT}$/$c_{\rm s}$). 

The statistical properties of the sample are summarised in Table\,\ref{table:average} including the mean values and the medians of the derived quantities, as well as the minimal and maximal values. 
The derived values of $T_{\rm ex}$, $T_{\rm rot}$, $T_{\rm kin}$, $N_{\rm tot}$, $\sigma_{\rm TH}$, $\sigma_{\rm NT}$, $c_{\rm s}$, $\mathcal{M}$ and $R_{\rm P}$ for individual Planck sources are listed separately in Table\,\ref{NH3_line} of Appendix\,\ref{appendixA}. 

\section{Results}
\label{sect:results}


\subsection{NH$_3$ detection rates}
\label{detection_rate}
Out of the 672 Planck Early Release Cold Cores (ECCs) that were surveyed with the Nanshan 26-m telescope for NH$_{3}$, 249 sources (37\%) were detected with NH$_{3}$\,(1,1) emission lines. Of these, 187 sources reveal NH$_{3}$\,(1,1) hyperfine structure, while 76 sources (11\%) also show NH$_{3}$\,(2,2) emission lines. Observed parameters and calculated parameters are given in Tables\,\ref{NH3_line}-\ref{NH_para} of Appendix\,\ref{appendixA}.

The highest angular resolution of the Planck survey is 5 arcmin, at a frequency of 857\,GHz. We plotted the distribution of observed sources and the NH$_{3}$ lines detected in this work as well as the detection rate of the NH$_{3}$\,(1,1) line as a function of the 857\,GHz aperture flux density S$_{857}$ of the Planck Sources in the \textit{lower left panel} of Fig.\,\ref{position}. We observe that sources with S$_{857}$ larger than 3\,Jy have a 100\% detection rate. Furthermore, the detection rate of the NH$_{3}$\,(1,1) line increases from 0 to 100 percent as the flux density S$_{857}$ rises from 1.5 to 4.6\,Jy. However, it should be noted that while the detection rates of the NH$_{3}$\,(1,1) line are 100 percent in the last bins, the number of sources within these bins is limited, with only 8, 6, 4, 3, 2, and 2 sources in total.

In the \textit{lower right panel} of Fig.\,\ref{position}, we observed an unexpectedly low detection rate of ammonia in regions with small kinematic distances. To understand this phenomenon, we performed a column density analysis of sources situated within a distance of less than 0.45 kpc. Utilizing hydrogen column density data from the \cite{2011A&A...536A..23P}, we found a difference in column densities between sources where ammonia was detected and those where it was not, within this near distance (less than 0.45 kpc). Specifically, sources with detected ammonia exhibited an average hydrogen column density of 4.32 $\times$ 10$^{21}$ cm$^{-2}$, while those without ammonia detection had an average hydrogen column density of 1.99 $\times$ 10$^{21}$ cm$^{-2}$. Moreover, it's important to consider that sources in these nearby regions may have relatively extended spatial distributions. This spatial extension implies that during our observations of ammonia point sources, we may inadvertently miss the real peaks of the Planck cold cores.

The top panel of Fig.\,\ref{position} presents the spatial distribution of the detected sources, which are mainly located in local star-forming regions such as Taurus and Orion and the galactic plane. This trend is shared by all of the ECCs and CO-selected cores in the sample \citep{2012ApJ...756...76W,2016ApJ...820...37Y}. As mentioned in Sect.\ref{Source}, we cross-matched our sample with sources observed by \cite{2022ApJS..258...17F}, owing to distinct criteria detailed in Section. \ref{sect:Introduction}. Consequently, there is a disparity in the distribution of sources across the galaxy between the two samples (see also Sect.\ref{Source}).

\subsection{Properties of the NH$_3$ emitting gas}
\label{mitting-gas}
\begin{figure*}[h]
\vspace*{0.2mm}
\centering
\includegraphics[width=0.98\textwidth]{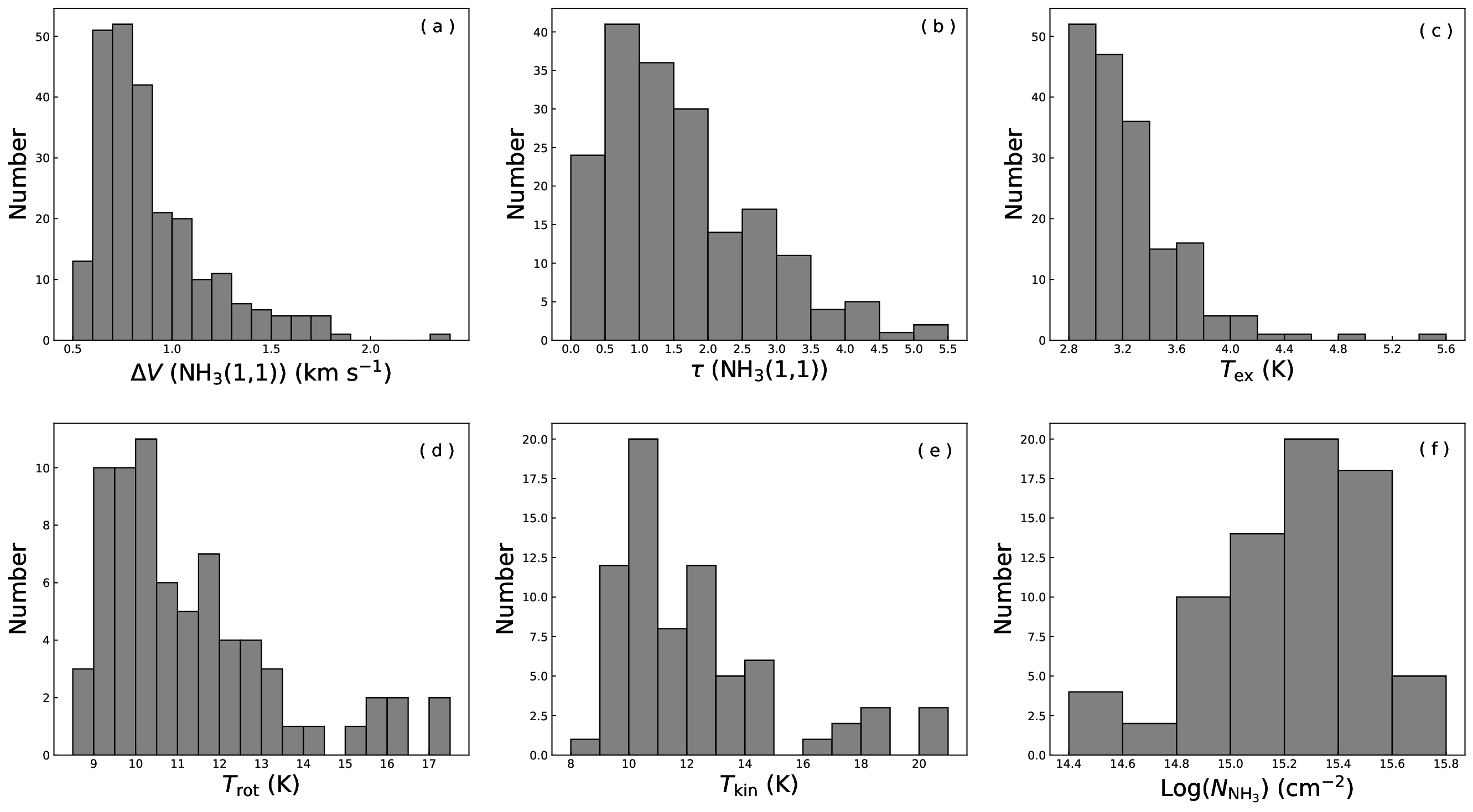}
\caption[]{Histograms of physical parameters derived from NH$_{3}$. (a) Intrinsic line widths of individual NH$_{3}$\,(1,1) hyperfine structure components with hyperfine structure and a peak line flux threshold of 3$\sigma$; (b) peak optical depths of the main group of hf components $\tau_{m}$ (1,1) for those positions with NH$_{3}$\,(1,1) hyperfine structure and S/N > 3 (these line widths and peak optical depths are derived from the GILDAS built-in “NH$_{3}$\,(1,1)” fitting method); (c) excitation temperature $T_{\rm ex}$; (d) rotational temperature $T_{\rm rot}$; (e) kinetic temperature $T_{\rm kin}$; (f) NH$_3$ column densities.}
\label{width}
\end{figure*}

Figure\,\ref{width} shows the statistics of observed and physical properties for the NH$_3$ detected sample. The top left panel shows the intrinsic line width of an individual hfs component, $\Delta V\,(\rm NH_{3}\,(1,1))$, which ranges approximately from 0.36 to 2.36\,km s$^{-1}$, with an average of 0.89 $\pm$ 0.29\,km s$^{-1}$, which suggests that the non-thermal turbulence of these sources is significant (errors correspond to the standard deviations of the mean throughout this article). The top middle panel of Figure\,\ref{width} shows the sample optical depths, which range from 0.1 to 5.3 with an average value of 1.6, implying that NH$_3$\,(1,1) lines in most of the detected PGCC cores are optically thick (see Table\,\ref{Tmean}). 

The top right panel shows the excitation temperature, $T_{\rm ex}$, which ranges approximately from 2.8 to 5.4\,K, with a mean and median of 3.2 and 3.1\,K, respectively. The bottom left panel of Figure\,\ref{width} shows the derived rotational temperature, which exhibits a range of 8.6 to 17.6\,K, with an average value of 11.4\,$\pm$\,2.2\,K. The median value of the rotational temperature is 10.7\,K, with a typical value of approximately 10\,K. Remarkably, 83\% of the sources exhibit values lying between 8 and 14\,K. The NH$_{3}$ kinetic temperature distribution, illustrated in the \textit{bottom middle panel} of Fig.\,\ref{width}, ranges from 8.9 to 20.7\,K, with an average of 12.3 $\pm$ 2.9\,K and a median value of 11.4\,K. The total NH$_3$ column density ranges from 0.36 to 6.07 $\times$ 10$^{15}$ cm$^{-2}$, with an average of 2.04 $\times$ 10$^{15}$ cm$^{-2}$. The total column densities of NH$_3$ are presented as a histogram in the \textit{bottom right panel} of Fig.\,\ref{width}, exhibiting a peak around 10$^{15.4}$ cm$^{-2}$ in our sample.

The thermal velocity dispersion of NH$_3$\,(1,1) lines detected at a $>$3$\sigma$ level shows a range of 0.07 to 0.10\,km\,s$^{-1}$, with an average of 0.08\,$\pm$\,0.01\,km\,s$^{-1}$ and a median of 0.07\,km\,s$^{-1}$. The non-thermal velocity dispersion of NH$_3$\,(1,1) for cores ranges from 0.30 to 1.09\,km\,s$^{-1}$, with an average of 0.55\,$\pm$\,0.18\,km\,s$^{-1}$ and a median of 0.49\,km\,s$^{-1}$. The thermal linewidth is significantly smaller than the non-thermal linewidth, which suggests that non-thermal motions dominate the dense gas in the PGCCs. The sound speed of the gas ranges from 0.18 to 0.27\,km\,s$^{-1}$, with an average of 0.21\,$\pm$\,0.02\,km\,s$^{-1}$ and median of 0.20\,km\,s$^{-1}$. The thermal to non-thermal pressure ratio in the gas traced by NH$_3$\,(1,1) ranges from 0.01 to 0.06, with an average of 0.02\,$\pm$\,0.01, and a median of 0.02. The Mach number ranges from 1.6 to 5.0 with an average of 2.7\,$\pm$\,0.8, and a median of 2.5.

\begin{table}
\caption{NH$_3$ parameters of our sample of PGCC cores}
\label{Tmean}
\centering
\small
\begin{tabular}
{lccccccc}
\hline\hline
Parameter  & Range & Mean &  Median \\
\hline
$\Delta$$V$(NH$_{3}$(1,1))    [km\,s$^{-1}$]         & 0.36--2.36  &   0.89  $\pm$ 0.29  & 0.83    \\  
$\tau$                                   & 0.1--5.3    &   1.6   $\pm$ 1.1   & 1.4     \\  
$T_{\rm ex}$      [K]                    & 2.8--5.4    &   3.2   $\pm$ 0.4   & 3.1     \\  
$T_{\rm rot}$     [K]                    & 8.6--17.6   &   11.4  $\pm$ 2.2   & 10.7    \\  
$T_{\rm kin}$     [K]                    & 8.9--20.7   &   12.3  $\pm$ 2.9   & 11.4    \\  
N$\rm _{NH_{3}}$  [$10^{15}$ cm$^{-2}$]  & 0.36--6.07  &  2.04   $\pm$ 1.2   & 1.83    \\  
$\sigma_{\rm TH}$ [km\,s$^{-1}$]         & 0.07--0.10  &   0.08  $\pm$ 0.01  & 0.07    \\  
$\sigma_{\rm NT}$ [km\,s$^{-1}$]         & 0.30--1.09  &   0.55  $\pm$ 0.18  & 0.49    \\  
$c_{\rm s}$       [\,km\,s$^{-1}$]       & 0.18--0.27  &   0.21  $\pm$ 0.02  & 0.20    \\  
$\mathcal{M}$                            & 1.6--5.0    &   2.7   $\pm$ 0.8   & 2.5     \\  
$R_{\rm P}$                              & 0.01--0.06   &   0.02 $\pm$ 0.01  & 0.02    \\  
$N\rm _{H_{2}}$($^{13}$CO) [$10^{22}$ cm$^{-2}$]&0.07--2.88&0.74 $\pm$ 0.42  & 0.63    \\ 
$\chi_{\rm \tiny NH_{3}}$ [$10^{-7}$]    & 0.3-9.7     &   2.7   $\pm$ 2.2   & 1.9     \\  
 $T_{\rm dust}$   [K]                    & 8.6-13.9    &   11.5  $\pm$ 1.07  & 11.6    \\

\hline  
\end{tabular}
\label{table:average}
\tablefoot{The dust temperature ($T_{\rm dust}$) values presented in this table correspond to a cross-matching of NH$_3$-detected 249 sources with the dust temperature data obtained from \cite{2016A&A...594A..28P}.}
\end{table}

\subsection{Thermal and non-thermal motions}
\label{sect-4-2}
The clumps are supported against their gravity by both thermal and non-thermal motions \citep{2003RMxAC..15..293C}. The former is a manifestation of the kinetic temperature within a clump, while the latter originates from star-forming activities such as infall motions and outflows that can broaden the non-thermal velocity dispersion. However, the quiescent nature of PGCCs suggests that these motions are not particularly vigorous. Therefore, turbulent motion is the primary contributor to the non-thermal motion of these cores. Ammonia is one of the few dense gas tracers that allows for the simultaneous computation of line width and kinetic temperature, as well as thermal and non-thermal line widths. Our data are particularly well-suited for investigating these two types of motion.

\begin{figure*}[t]
\centering
\includegraphics[width=0.65\textwidth]{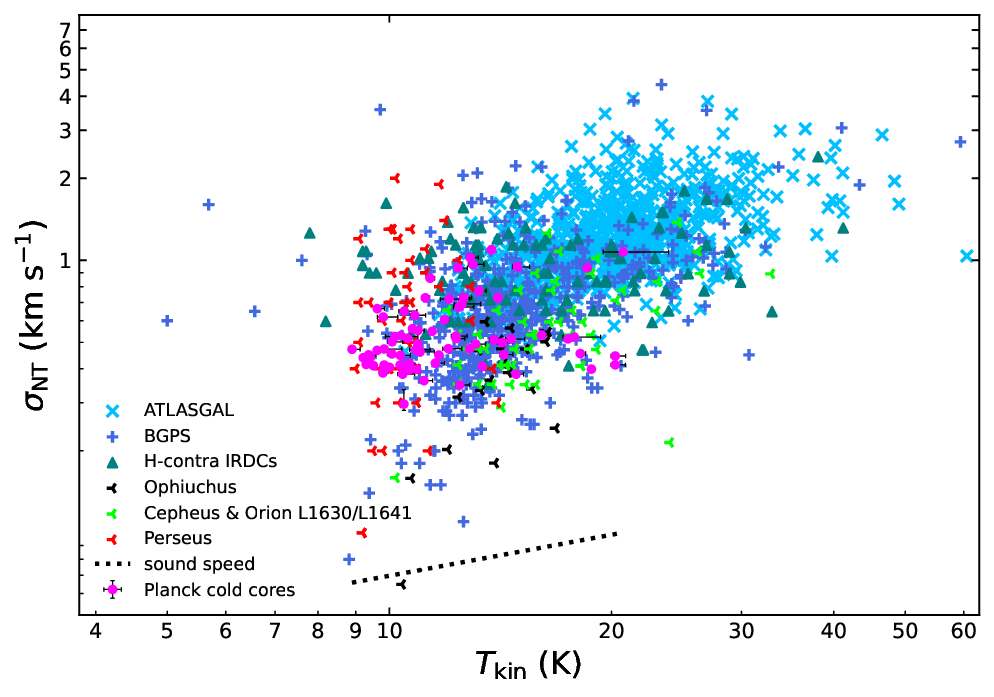}
\caption[]{Left: Non-thermal velocity dispersion $\sigma_{\rm NT}$ vs. gas kinetic temperature derived from NH$_{3}$\,(1,1). The black dotted line at the bottom represents the thermal sound speed (\textit{left}).}
\label{sigmav}
\end{figure*}

When turbulent energy is converted into heat, a correlation is expected to exist between the kinetic temperature and linewidth \citep{1985A&A...142..381G, mol96, 2013A&A...550A.135A, 2016A&A...586A..50G, 2016A&A...595A..94I, 2017A&A...598A..30T, 2018A&A...609A..16T, 2018A&A...611A...6T, 2021A&A...655A..12T}. In this study, we investigate the existence of a correlation between temperature and turbulence in PGCCs traced by the dense gas. To achieve this, we utilized the NH$_3$\,(1,1) and (2,2) line ratio derived kinetic temperature and NH$_3$ non-thermal velocity dispersion ($\sigma_{\rm NT}$) as a decent approximation proxy for turbulence. The non-thermal velocity dispersions versus the kinetic temperature for the sources are shown in Fig.\,\ref{sigmav}, with the dotted line representing the thermal sound speed. Our results indicate a weak correlation between non-thermal velocity dispersion and kinetic temperature, suggesting that turbulent heating may contribute to gas temperature in these cold cores.

\subsection{Associated stellar objects}

\begin{table*}
\caption{Numbers of associated stellar objects}
\label{associate}
\tiny
\centering
\begin{tabular}{|p{2.75cm}<{\centering} p{1cm}<{\centering} | p{1.4cm}<{\centering} p{1.55cm}<{\centering} p{1.5cm}<{\centering}| p{1.4cm}<{\centering} p{1.7cm}<{\centering} p{2cm}<{\centering}|}
\hline\hline
\centering
Types of matched cores & Matching criteria& \multicolumn{3}{c}{Number of associated stellar objects}& \multicolumn{3}{|c|}{Number of cores containing stellar objects} \\
\hline
 && with IR objects& with YSOs& with YSO candidates & contain IR objects& contain YSOs& contain YSO candidates\\
 \hline
672 observed cores       & $\sim$2$\arcmin$& 131 ($\sim$19\%)& 57 ($\sim$8\%)& 58 ($\sim$8\%)& 112 ($\sim$17\%)& 29 ($\sim$4\%)& 23 ($\sim$3\%)\\ 
                         
249 NH$_3$ detected cores & $\sim$2$\arcmin$& 66 ($\sim$27\%)& 48 ($\sim$19\%)&  52 ($\sim$21\%)& 53 ($\sim$21\%)& 23 ($\sim$9\%)& 17 ($\sim$7\%)\\
                          
\hline
\hline 
\end{tabular}
\tablefoot{Matching criteria refer to associated objects within the $\sim$2$\arcmin$ ammonia beam size.}
\end{table*}
\begin{table*}
\caption{Matching results by types of stellar objects}
\label{type1}
\tiny
\centering
\begin{tabular}{|p{2.75cm}<{\centering} p{1cm}<{\centering} | p{1.4cm}<{\centering} p{1.1cm}<{\centering} p{1.1cm}<{\centering} |p{1.65cm}<{\centering} p{1.35cm}<{\centering} p{1.4cm}<{\centering} p{2cm}<{\centering}|}
\hline\hline
\centering
Types of matched cores  & Matching criteria& \multicolumn{3}{c}{Single-source matches} &\multicolumn{4}{|c|}{Multi-source matches}\\
\hline
  && IR  & YSO &  YSO candidate & YSOs \& YSO candidate & IR \& YSOs &IR \& YSO candidate& IR\&YSOs\&YSO candidate \\
\hline
672 observed cores          & $\sim$2$\arcmin$& 102 ($\sim$15\%)& 18 ($\sim$3\%)& 16 ($\sim$2\%)& 4 ($\sim$1\%)& 7 ($\sim$1\%)&  3 ($\sim$1\%)& \,\\
                            
249 NH$_3$ detected cores   & $\sim$2$\arcmin$& 45 ($\sim$18\%)& 13 ($\sim$5\%)& 9 ($\sim$4\%)& 5 ($\sim$2\%)& 5 ($\sim$2\%)& 3 ($\sim$1\%)& \,\\

\hline
\hline 
\end{tabular}
\tablefoot{"Single-source matches" are sources that are exclusively associated with either an IR source or with one specific type of stellar object. "Multi-source matches" refer to sources that are associated with multiple types.}
\end{table*}

The associated objects of the cores are crucial for our understanding of their environment and evolutionary stages. Therefore, we investigated the objects associated with the cores, including infrared (IR) objects, young stellar objects (YSOs), and young stellar object candidates (YSO candidates). We obtained the stellar objects from the Simbad website. Statistical properties are given in Tables\,\ref{associate}-\ref{type1}. YSOs are distinguished by their manifestation of not only infrared excess but also additional spectral features, providing evidence of active star formation processes \citep{2004ApJS..154..363A, 2004ApJS..154..374G}. Conversely, YSO candidates represent potential YSOs necessitating further scrutiny and validation.

In our study, we adopted the ammonia beam size ($\sim$2$\arcmin$) as a matching criterion. Among the 672 cores observed, our analysis revealed the presence of 131 ($\sim$19\%) IR objects, 57 ($\sim$ 8\%) YSOs and 58 ($\sim$ 8\%) YSO candidates. The rare association may indicate low star formation activity \citep{2009ApJ...703...52L,2010A&A...512A..67L}. However, it cannot be entirely excluded that the associated objects originate from different gas clumps located along the line-of-sight \citep{2006A&A...450..607W}. It is important to note that certain cores in our sample may encompass multiple IR sources, as well as multiple YSOs or YSO candidates. Notably, the majority of these IR objects correspond to single-point sources detected by the Infrared Astronomical Satellite (IRAS). Overall, our findings indicate that 149 ($\sim$22\%) cores have different types of associated objects within a beam size of  $\sim$2$\arcmin$. Additionally, we find that 112 ($\sim$17\%) cores contain at least one IR object, 29 ($\sim$4\%) cores have YSOs, and 23 ($\sim$3\%) cores contain YSO candidates. Among these, 102 sources only associate with IR objects, 18 cores only associate with YSOs and 16 cores only associate with YSO candidates. 4 cores are associated with both YSOs and YSO candidates, 7 cores with both YSO and IR objects, 3 cores with both IR objects and YSO candidates.

Considering the 249 cores where NH$_3$\,(1,1) emission was detected, we find 66 ($\sim$27\%) IR objects, 48 ($\sim$19\%)YSOs, and 52 ($\sim$21\%) YSO candidate sources associated with these cores using the aforementioned $\sim$2$\arcmin$ criterion. In general, 80 ($\sim$32\%) cores have different types of associated objects. Additionally, we find that 53 ($\sim$21\%) cores contain at least one IR object, 23 ($\sim$9\%) cores host YSOs, and 17 ($\sim$7\%) cores harbor YSO candidates. 45 sources only associate with IR objects, 13 cores only associate with YSOs and 9 cores only associate with YSO candidates. 5 cores are associated with both YSOs and YSO candidates, 5 cores with both YSOs and IR objects, and 3 cores with both IR objects and YSO candidates.

Based on preliminary statistical analysis, the proportion of sources with NH$_3$ emission matching IR sources and stellar objects is higher than the overall proportion of the total sample matching IR sources and YSOs. The ammonia detection rate of sources matching young stellar objects or their candidates in the 672 observed cores is 66\%, while the ammonia detection rate of sources matching IR objects is 45\%. The ammonia detection rate is higher in sources with matching stellar objects. Not all sources with matched stellar objects and IR objects have been detected in NH$_3$\,(1,1). There can be several reasons for this. Firstly, distance plays a significant role. If the YSOs or IR objects are located at a large distance, the faint emission signal from ammonia may be too weak to be detected. Secondly, the spatial resolution of the observing instrument can also be a factor. Other factors include line-of-sight effects and environmental conditions, where the physical environment surrounding the YSOs may influence the presence of ammonia molecules. The inescapable issue at hand is that our matching could potentially be limited to line-of-sight coincidence, lacking the necessary distance information for spatial alignment.

\subsection{NH$_3$ abundances}
\label{sect-4-3}
The column densities of ammonia in our sample range from 0.36 $\sim$ 6.07 × 10$^{15}$ cm$^{-2}$, with a mean value of approximately 2.04 $\times$ 10$^{15}$ cm$^{-2}$ and a median value of about 1.83 × 10$^{15}$ cm$^{-2}$ (Table\,\ref{Tmean}). We utilized $^{13}$CO to derive the H$_{2}$ column densities as described in Appendix\,\ref{appendB}. The H$_{2}$ column densities $N\rm _{H_{2}}$($^{13}$CO) range from 7.0 $\times$ 10$^{20}$ to 2.88 $\times$ 10$^{22}$ cm$^{-2}$, with an average of 7.4 $\times$ 10$^{21}$ cm$^{-2}$. The column densities of NH$_{3}$ denoted as $N_{\rm NH_{3}}$, were compared with the column densities of H$_{2}$, denoted as $N_{\rm H_{2}}$, derived from $^{13}$CO to determine the fractional abundance of ammonia in each source. The fractional abundance, $\chi_{\rm NH_{3}}$, is defined as the ratio of $N_{\rm NH_{3}}$ to $N_{\rm H_{2}}$($^{13}$CO). The ammonia abundances in the sources range from 0.3 to 9.7 $\times$ 10$^{-7}$, with an average of 2.7 $\times$ 10$^{-7}$. 

\begin{figure*}[th!]
\centering
\includegraphics[width=0.8\textwidth]{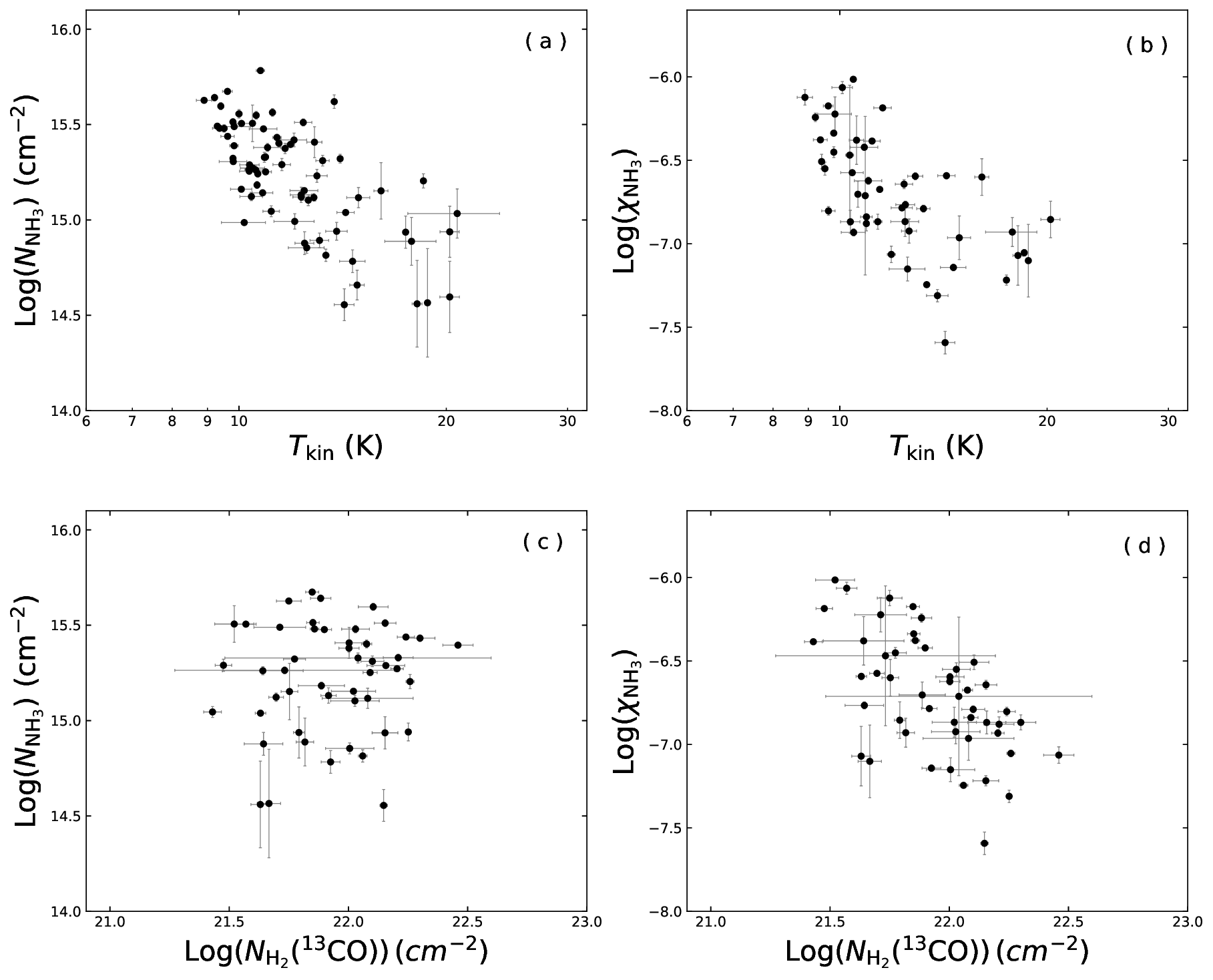}
\caption[]{
(a,b) Column densities and fractional abundances $N_{\rm NH_{3}}$/$N_{\rm H_{2}}$($^{13}$CO) versus kinetic temperature $T_{\rm kin}$ and (c,d) vs. the column density of H$_{\rm 2}$ derived from $^{13}$CO ($N_{\rm H_{2}}$($^{13}$CO)).}
\label{Nnh3}
\end{figure*}

Fig.\,\ref{Nnh3} illustrates the column densities of NH$_{3}$ and its abundances, $\chi_{\rm NH_{3}}$, as a function of H$_{2}$ column density and kinetic temperature $T_{\rm kin}$. It is observed that the NH$_{3}$ column densities and fractional abundances are inversely proportional to the kinetic temperatures. Moreover, there is a trend of decreasing fractional NH$_{3}$ abundance, $\chi_{\rm NH_{3}}$, with increasing $N_{\rm H_{2}}$.

\section{Discussion}
\label{discussion}
\subsection{Column densities and abundances}

The ammonia column densities in our results are consistent with the findings of PGCC selected sources from SCUBA-2, whose column densities range from 0.4 to 1.5 $\times$ 10$^{16}$ cm$^{-2}$, with a mean value of approximately 1.3 $\times$ 10$^{15}$ cm$^{-2}$ \citep{2022ApJS..258...17F}. The ammonia column densities we obtained in our observations are also consistent with those reported in most star-forming environments \citep{dun11, wie12, cyg13}.

The average value of $N\rm _{H_{2}}$($^{13}$CO), 7.4 $\times$ 10$^{21}$ cm$^{-2}$, is consistent with the results obtained from the Planck cold cores observed by PMO \citep{2012ApJ...756...76W, 2013ApJS..209...37M, 2016ApJS..224...43Z}. In the SCOPE-2 follow-up observations, the H$_{2}$ column densities of 97 Planck cold cores range from 10$^{22}$ to 10$^{24}$ cm$^{-2}$ \citep{2022ApJS..258...17F}, which are higher than the column densities obtained in our work. This may be attributed to the fact that the SCOPE-2 survey does not include sources with column densities $N_{\rm H_{2}}$ < 2 $\times$10$^{21}$ cm$^{-2}$ \citep{2019MNRAS.485.2895E}.

Previous observations of NH$_{3}$ fractional abundance in high-mass star-forming clumps suggest a median value of 2.7 $\times$ 10$^{-8}$ \citep{2015MNRAS.452.4029U}. Additionally, \cite{dun11}, \cite{wie12}, and \cite{2019MNRAS.483.5355M} obtained average abundance values of 1.2 $\times$ 10$^{-7}$, 4.6 $\times$ 10$^{-8}$, and 1.5 $\times$ 10$^{-8}$, respectively, in clumps of the Bolocam Galactic Plane Survey (BGPS), the APEX Telescope Large Area Survey of the GALaxy (ATLASGAL), and the Hi-GAL survey. Furthermore, fractional abundances of 4 $\times$ 10$^{-8}$, 8 $\times$ 10$^{-7}$ and 1 $\times$ 10$^{-8}$ were derived for infrared dark clouds by \cite{2006A&A...450..569P}, \cite{2011ApJ...736..163R}, and \cite{chi13}, respectively. The average NH$_{3}$ abundance derived from PGCCs in this study is one order of magnitude greater than that of MSF regions and IRDCs and exhibits a narrow distribution similar to that of IRDCs. The elevated NH$_{3}$ fractional abundance in PGCCs relative to other star-forming regions is likely attributable to the earlier phase of cloud evolution of PGCCs compared to Massive Star-forming (MSF) regions or even IRDCs. Ammonia is more abundant in cold, dense cores than most other molecules \citep{1997ApJ...486..316B}. The average NH$_{3}$ abundances obtained in our study are consistent with those predicted for low-mass pre-protostellar cores by chemical models \citep{1997ApJ...486..316B}.

The NH$_{3}$ column densities and fractional abundances are inversely proportional to the kinetic temperatures (see in Fig.\,\ref{Nnh3} a, b), which is consistent with the findings of previous studies on infrared dark clouds \citep{chi13}. The NH$_{3}$ column densities in all cores tend to increase with $N_{\rm H_{2}}$, as shown in Fig.\,\ref{Nnh3} (c), following the general trend that the NH$_{3}$ emission follows the dust continuum emission in the cores. This is in agreement with the results reported by \cite{2011ApJ...736..163R}, who found that IRDCs with no evidence of embedded star formation activity exhibit a strong correlation between ammonia and H$_{2}$ column density. A similar trend is observed in the cold cores in our observation.

\begin{table*}
\caption{Fundamental information on different NH$_3$ samples}
\label{type}
\centering
\small
\begin{tabular}{lccccccc}
\hline\hline
Sources   & Reference & Type& NH$_3$ source number \\
\hline
BGPS                    & \cite{dun11} & High-mass star-forming clumps,                       & 456   \\                                         
                        &              & at various evolutionary stages                               \\                                         
ATLASGAL                & \cite{wie12} & High-mass star-forming clumps,                       &752    \\                                         
                        &              & at various evolutionary stages                               \\                                         
UC H{\sc ii} Cand       & \cite{mol96} & Percursors of ultraconpact H{\sc ii} regions          &163    \\          
High contrast IRDCs          & \cite{chi13} & Capable of harbouring high-mass star-forming regions &109    \\                               
                        &              & Early evolutionary stages                                    \\                                         
High IR extinction clouds  & \cite{ryg10} & Cold and compact high extinction clouds              &54     \\                                         
EGOs                    & \cite{cyg13} & Northern extended green objects (EGOs)               &94     \\                                         
Ophiuchus               & \cite{fri09} & Studies of isolated starless clumps                  &21     \\                                         
Perseus                 & \cite{sch09} & Starless cores                                       &49     \\                                         
Cepheus \& Orion L1630 / L1641  & \cite{har93,har91} & Warm, massive clumps                                 &51     \\

\hline 
\end{tabular}
\end{table*}

\subsection{Comparative analysis of thermal and non-thermal motions}

To obtain comprehensive statistics on PGCCs, we conducted a comparative analysis of the non-thermal velocity dispersions and kinetic temperatures. Studies encompass a diverse range of samples, including high-mass star-forming clumps at various evolutionary stages from ATLASGAL and BGPS surveys \citep{dun11, wie12}, high contrast Infrared Dark Clouds (IRDCs) \citep{chi13}, as well as starless cores in the Perseus region \citep{row08, sch09}, and dense cores in the Ophiuchus, Cepheus, and Orion L1630 and L1641 molecular clouds \citep{har91, har93, fri09}. This comprehensive approach allows us to gain valuable insights into the physical properties of these objects and their evolutionary characteristics. In Table\,\ref{type}, we have compiled a list containing the fundamental information of the sources used for comparison in this study.

As seen in Fig.\,\ref{sigmav}, unlike high-mass star-forming clumps and dense cores in star-forming regions \citep{dun11, wie12}, the relationship between non-thermal velocity dispersion and kinetic temperatures in PGCCs is much weaker, which is similar to or even less than for infrared dark clouds \citep{chi13}. Additionally, the distribution of non-thermal velocity dispersion and kinetic temperatures of Planck cold cores is narrower when compared to the observations of the other samples. Spearman’s rank correlation coefficients ($\rho$) for non-thermal velocity dispersion versus kinetic temperatures in different samples and the resultant $\rho$-values are presented in Table \ref{Spearman-test} of Appendix \ref{appendixA}. The table presents the obtained $\rho$-values along with their corresponding P-values. The analysis of this table reveals varying degrees of correlation between non-thermal velocity dispersion and kinetic temperature across different samples. Particularly, a conspicuous positive correlation is discernible in the ATLASGAL, BGPS, Ophiuchus, Cepheus, and Orion L1630/L1641 samples, underscored by small P-values. However, in the High-contrast IRDCs, the correlation is relatively weak, with a Spearman correlation coefficient of 0.07 and a p-value of 0.45, suggesting a weak and non-significant linear relationship between these two variables. The Perseus sample also manifests a relatively weak correlation, featuring a Spearman correlation coefficient of 0.2 and a p-value of 0.16, implying a modest positive correlation trend that lacks statistical significance. In contrast, our Planck cold cores sample displays a pronounced positive correlation between non-thermal velocity dispersion and kinetic temperature. With a Spearman correlation coefficient of 0.33 and a p-value of 4.00 $\times$ $10^{-3}$.

Furthermore, gas pressure ratios derived from our results closely align with those obtained by \cite{2015MNRAS.452.4029U}, who conducted a study on 66 massive young stellar objects and compact H{\sc ii} regions from the RMS survey (Red MSX Source survey, \citealp{2013ApJS..208...11L}). \cite{2015MNRAS.452.4029U} obtained gas pressure ratios ($R_{\rm p}$) of 0.01-0.02 for both massive star-forming and quiescent clumps. In contrast, \cite{2003ApJ...586..286L} reported significantly higher values of $R_{\rm p}$ = 4-5 for low-mass star-forming cores, such as Barnard 68, based on C$^{18}$O and C$^{34}$S data. Our sample, which includes non-thermal movements that significantly contribute to the kinetic energy balance, more closely aligns with samples of massive clumps compared to others.

\subsection{Correlation between NH$_3$ and other molecular species}

\begin{figure*}[h]
\vspace*{0.2mm}
\centering
\includegraphics[width=0.89\textwidth]{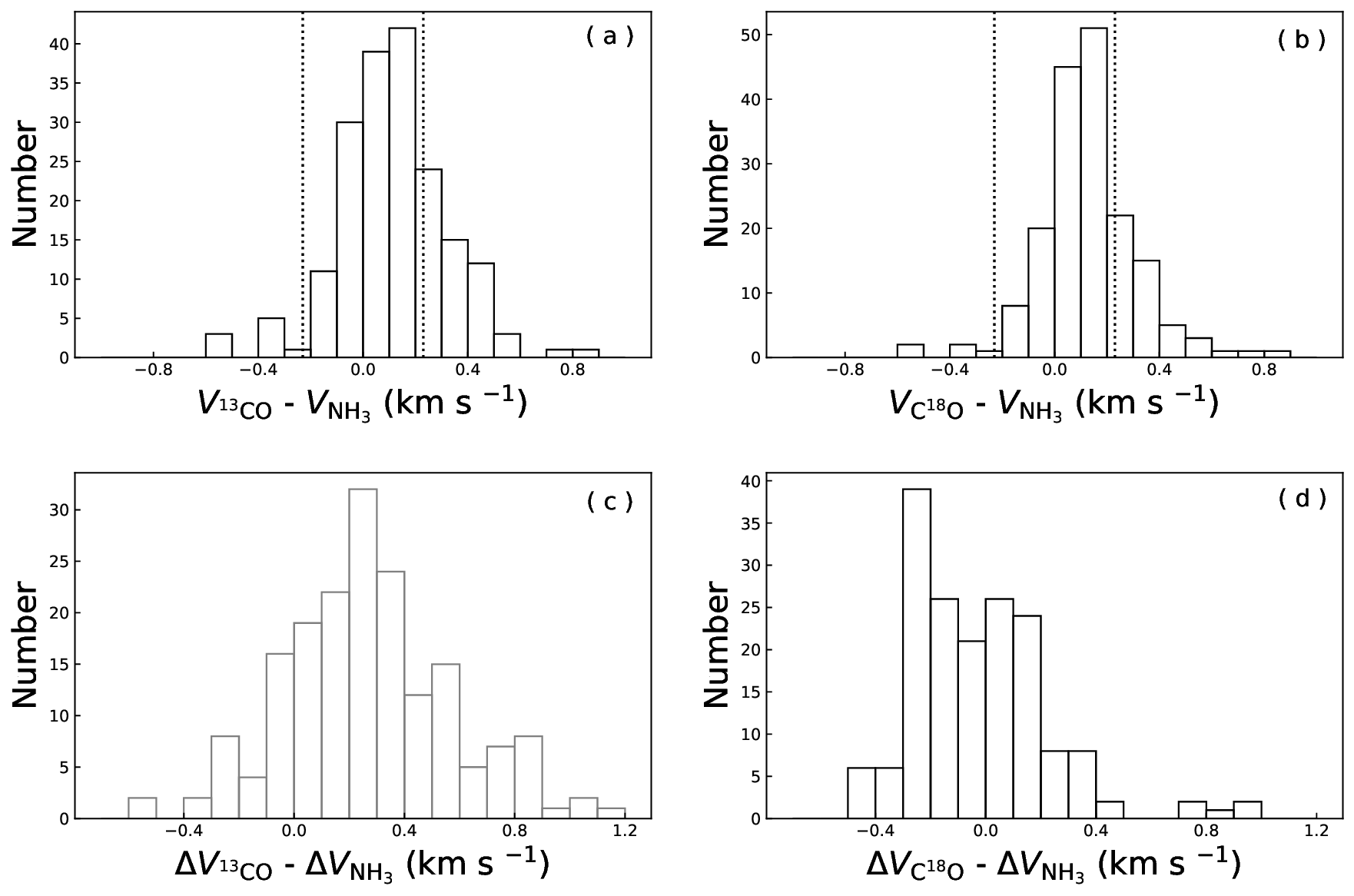}
\caption[]{Distributions between differences of line-center velocities (a, b) and line widths (c, d) of $^{13}$CO, C$^{18}$O and NH$_{3}$. The average sound speed is indicated as dotted lines in (a, b).}
\label{kinimatic_v}
\end{figure*}

In their survey of 674 Planck cold clumps, \cite{2012ApJ...756...76W} discovered that the velocities of transitions from CO and its isotopologues were highly consistent with one another. We plot the absolute value of the difference between $J$=1-0 transitions of $^{13}$CO and C$^{18}$O and the NH$_3$(1,1) line-center velocities from Gaussian fits in Fig.\,\ref{kinimatic_v} (a, b). When multiple $^{13}$CO velocities were detected along the line of sight, we assumed that the velocity component closest to the NH$_{3}$ velocity is associated with the core \citep{2007ApJ...668.1042K}.

The minute discrepancy in centroid velocity between NH$_{3}$ and $^{13}$CO ($\sim$0.17\,km\,s$^{-1}$), and between NH$_{3}$ and C$^{18}$O ($\sim$0.18\,km\,s$^{-1}$), can be compared to the distinction in line-center velocities of N$_{2}$H$^{+}$ and C$^{18}$O in low-mass, starless, and protostellar cores \cite{2004ApJ...614..194W, 2007ApJ...668.1042K}.
Fig.\,\ref{kinimatic_v} (a, b) illustrate that the majority of cores (64\% for the difference between $^{13}$CO NH$_3$(1,1) line-center velocities and 70\% for the difference between C$^{18}$O and NH$_3$(1,1) line-center velocities) exhibit differences smaller than the sound speed (dotted lines), and the remaining cores have differences that are not much larger. This phenomenon was investigated in the Perseus molecular cloud by \cite{2007ApJ...668.1042K}, who found that the relative motions of dense N$_{2}$H$^{+}$ cores and their envelopes (measured in C$^{18}$O) generally exhibit velocity differences lower than thermal motions in the majority of cases ($\sim$ 90\%). This is also consistent with the findings of \cite{2004ApJ...614..194W}, who measured a core-to-envelope velocity difference that exceeded the sound speed in only 3\% of their sample. These samples consisted of isolated low-mass cores. Therefore, these comparisons suggest that the velocity differences between low and high-density tracers in the Planck cores observed by NH$_{3}$ are similar to those in low-mass, starless, or prestellar cores. In the survey of CO-selected cores in Planck cold clumps, \cite{2016ApJ...820...37Y} found that the central velocity differences of transitions from $^{13}$CO, HCO$^{+}$ and HCN are strikingly consistent with each other. The central velocities (obtained by fitting Gaussian profiles) between $^{13}$CO and HCO$^{+}$ ($V_{^{13}\rm CO}$ - $V_{\rm HCO^{+}}$ ), $^{13}$CO and HCN ($V_{^{13}\rm CO}$ - $V_{\rm HCN}$), HCO$^{+}$ and HCN ($V_{\rm HCO^{+}}$ - $V_{\rm HCN}$) in their study have mean values of 0.006, 0.05, and 0.002\,km\,s$^{-1}$, respectively. The smaller deviations from \cite{2016ApJ...820...37Y} may indicate that $^{13}$CO, HCO$^{+}$, and HCN are better coupled with each other. It is important to note that the channel spacing employed in their study was 0.21 km s$^{-1}$. In this work, the mean value of the velocity difference among the central velocities of $^{13}$CO and NH$_{3}$ is 0.17\,km\,s$^{-1}$, with channel spacing of 0.16 km s$^{-1}$. It is worth noting that the relatively low-velocity resolution might impact the interpretation of the small velocity differences observed in both cases, To address this, future observations with higher resolution would be beneficial for a more accurate comparison. By improving the velocity resolution, we can measure and compare the central velocity differences between different molecular lines more accurately, providing deeper insights. Such observations would help further validate and explain the consistency among these molecular transitions and enhance our understanding of the physical processes occurring within Planck cold clumps. Therefore, future high-resolution observations will be would help to further validate this matter.

The mean intrinsic line width of the NH$_{3}$\,(1,1) lines obtained from GAUSS fit is approximately 0.89\,km\,s$^{-1}$, which is comparable to that of C$^{18}$O $J$=1-0 (0.8\,km\,s$^{-1}$) in \cite{2012ApJ...756...76W}, C$_{2}$H $N$= 1-0 (1.0\,km\,s$^{-1}$) reported in \cite{2019A&A...622A..32L}, and HCN $J$ = 1-0 (1.06\,km\,s$^{-1}$) in \cite{2016ApJ...820...37Y}. The differences in line widths among different molecular tracers may result from turbulence at different scales. Fig.\,\ref{kinimatic_v} (c, d) show the distributions of the differences in line widths between $^{13}$CO and C$^{18}$O with NH$_{3}$. The mean widths of NH$_{3}$ and $^{13}$CO are nearly identical, with an average difference of $\sim$0.14\,km\,s$^{-1}$. In contrast to high-mass star-forming clumps, where the average difference between NH$_{3}$ and $^{13}$CO is 4.3\,km\,s$^{-1}$ \citep{wie12}, the intrinsic line widths of NH$_{3}$, $^{13}$CO and C$^{18}$O are approximately equal in our sample. This finding is consistent with the notion that these PGCCs are quiescent, as the majority of them appear to be transitioning from clouds to dense cores \citep{2012ApJ...756...76W}.

\subsection{A comparison with different star formation samples}
\label{sect-4-5}
To investigate the physical conditions and assess the potential for star formation within the cold clumps, we conducted a comparative analysis of various NH$_3$ molecular line surveys targeting different types of celestial objects. This analysis involved a comprehensive examination of line widths, kinetic temperatures, NH$_3$ column densities, as well as column densities of H$_{2}$, and fractional abundances. In this particular study, to conduct a more comprehensive comparative analysis, we have incorporated not only the samples mentioned in Sec.\,\ref{sect-4-2}, but also included additional statistical samples, ultra-compact H{\sc ii} (UC H{\sc ii}) regions or UC H{\sc ii} region candidates \citep{mol96}, High Infrared (IR) extinction clouds \citep{ryg10} and Extended Green Objects (EGOs) \citep{cyg13}. The fundamental information regarding these sources has been presented in Table\,\ref{type}. It is crucial to emphasize that not all papers provide data for every parameter. In certain studies, some parameters were not reported. Therefore, in the comparison involving these parameters, we opted to consider only those papers that presented available data.

\begin{figure*}[t]
\centering
\includegraphics[width=0.99\textwidth]{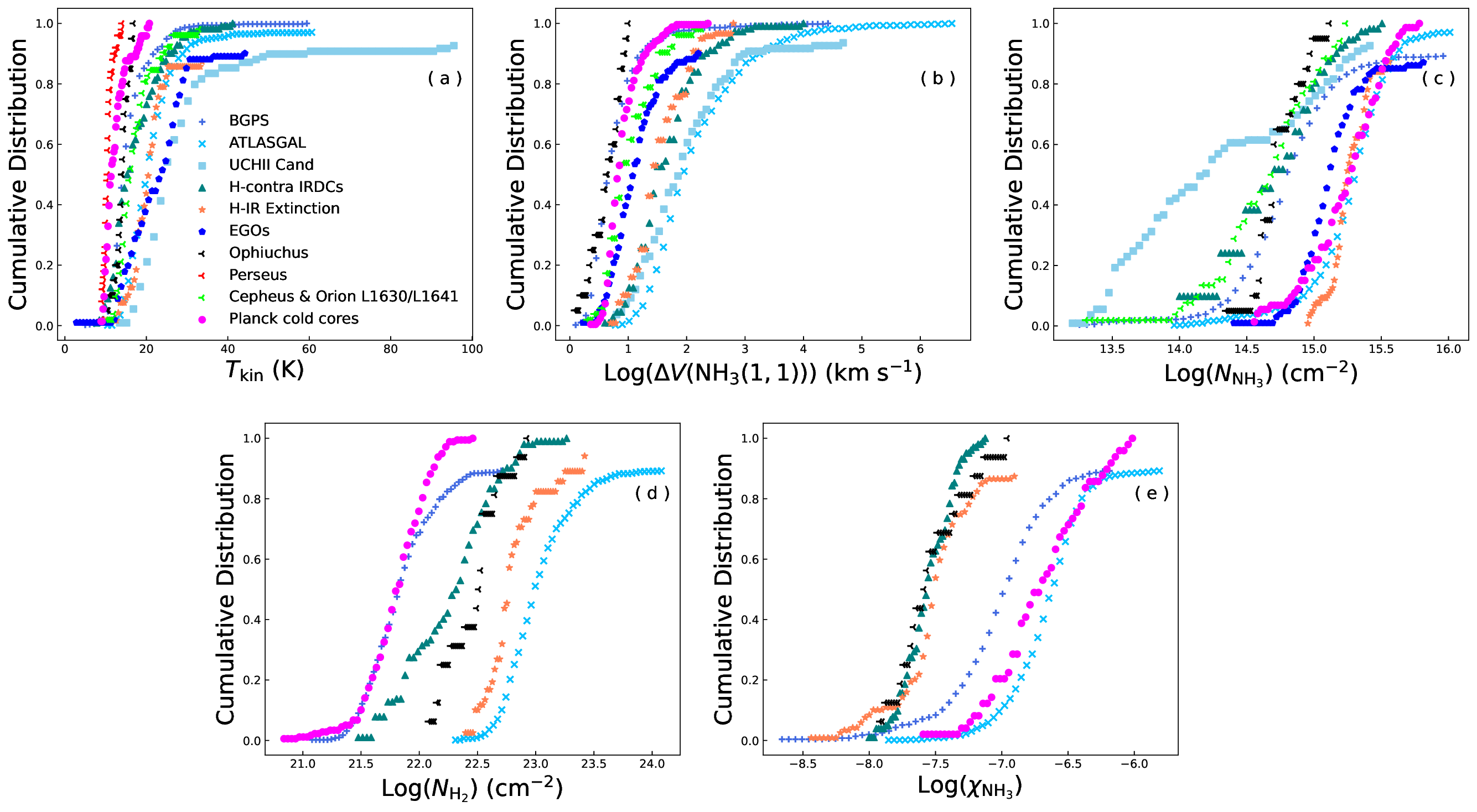}
\caption[]{Comparisons for the cumulative distribution of the line widths of NH$_3$\,(1,1), kinetic temperatures, column densities, and abundances of different star formation samples.}
\label{compare}
\end{figure*}

The kinetic temperatures of the cores range from 8.9 to 20.7\,K, with an average of 12.3 $\pm$ 2.9\,K and a median value of 11.4\,K (as mentioned in Sect.\,\ref{mitting-gas}),
and after cross-matching our NH$_3$ detected 249 sources with PGCCs from \cite{2016A&A...594A..28P}, we have observed that the dust temperature ($T_{\rm dust}$) of these 249 sources falls within the range of 8.6 to 13.9\,K, with an average value of 11.5\,K. The similarity in temperature ranges between $T_{\rm kin}$ and $T_{\rm dust}$ indicates an efficient coupling of gas and dust. The cumulative distribution of the kinetic temperature of NH$_3$ for different samples is presented in Fig.\,\ref{compare} (a). The results indicate that the UC H{\sc ii} candidates exhibit the highest kinetic temperatures \citep{mol96}, while the Planck cold cores show lower temperatures compared to other classes of sources. The temperature distribution of the samples is distinct and follows the expected evolutionary sequence. Hence, the kinetic temperature can serve as a reliable indicator for distinguishing between different evolutionary stages.

Fig.\,\ref{compare} (b) depicts the cumulative distribution of NH$_3$\,(1,1) line widths. Our analysis indicates that the line widths of ATLASGAL sources are the largest. The changes in the ATLASGAL and UC H{\sc ii} regions flatten off when the line width exceeds $\sim$2.5\,km\,s$^{-1}$, and their slopes are nearly identical when the line width is smaller than 2.5\,km\,s$^{-1}$. The variation of the cumulative fraction function of line width for the Planck cold cores is narrower when compared to other star-forming samples \citep{dun11, wie12, chi13}, and exhibits a similar cumulative distribution figure to those found in the dense cores of regions like Cepheus, and Orion L1630 and L1641, and Ophiuchus, as reported in \cite{har91, har93} and \cite{fri09}. However, the average linewidth in the Planck cold cores is slightly larger than that observed in Ophiuchus (0.62 km s$^{-1}$). Notably, the higher line width values observed in the Planck cold cores, in comparison to the dense cores in Ophiuchus, may imply that they are in more advanced and warmer stages of evolution. However, it is worth highlighting that the line width remains relatively small across the various samples under examination. Fig.\,\ref{compare} (c), (d), and (e) presents the cumulative distribution of column densities for NH$_3$, H$_{2}$, and fractional ammonia abundances for various samples. The ammonia column densities of cold cores are generally higher than those of other samples and are even comparable to those of ATLASGAL and EGO sources. Conversely, the cumulative distribution for Planck cold clumps exhibits the smallest H$_{2}$ column density range, resulting in high ammonia abundances. Despite the relatively high NH$_3$ column density observed in Planck cold cores, comprehensive consideration of its lower temperature, linewidth, and other parameters still leads to the conclusion that Planck sources are in an early stage of evolution.

To quantitatively compare the distributions, we conducted two-sample Kolmogorov-Smirnov (K-S) tests using the  \href{https://docs.scipy.org/doc/scipy/reference/generated/scipy.stats.ks\_2samp.html}{scipy.stats.ks\_2samp} procedure in the scipy package. This statistical test assesses the dissimilarity between two samples, with the K-S statistic serving as a measure of this dissimilarity, ranging from 0 to 1. A K-S statistic of 0 implies that there is no significant difference, suggesting the two samples may be from the same distribution. Conversely, a higher K-S statistic suggests a lower likelihood of the samples originating from the same distribution. Additionally, we computed P-values for each pair of samples, representing the probability that the samples share the same distribution. The outcomes of these tests are summarized in Table \ref{ks-test} of Appendix \ref{appendixA}.
In summary, the K-S test results for all examined parameters, including kinetic temperature, line width, ammonia column density, hydrogen molecule column density, and ammonia abundance, consistently yielded P-values below 0.05. These findings strongly imply substantial disparities between the Planck cold cores and the comparison samples in the examined parameters, which may reflect variations in their physical properties or the influence of distinct environmental conditions. Nevertheless, it is noteworthy that, in the distributions of kinetic temperature and linewidth, the Perseus, Cepheus, and Orion L1630/L1641 samples exhibit smaller KS statistic values. Moreover, their P-values are comparatively larger than those of other samples. In contrast, there is no pronounced similarity in the distributions of ammonia column density and hydrogen molecule column density among these samples. Additionally, in the abundance distribution, Planck displays relatively smaller KS values and larger P-values when compared to BGPS and ATLASGAL samples.

\section{Summary}
\label{sect:summary}
We used the Nanshan 26-m radio telescope to perform single-pointing observations of NH$_{3}$\,(1,1) and (2,2) inversion transitions towards 672 Planck sources. The main results of this study are as follows:
\begin{enumerate}
\item
Among the observed 672 Planck sources, 249 (37\%) were detected. Among these detections, 187 cores exhibit NH$_{3}$\,(1,1) hyperfine structure while 76 (11\%) cores also show corresponding NH$_{3}$\,(2,2) emission lines. The observed sources are mainly located in local star-forming regions. The detection rate of NH$_{3}$ is positively correlated with the continuum emission fluxes of Planck sources at a frequency of 857\,GHz, increasing as the 857\,GHz flux density increases. 

\item
Among the observed 672 sources, $\sim$22\% have associated stellar and IR objects within the beam size ($\sim$2$\arcmin$). This may indicate low star formation activity of the cores in our sample and the ammonia detection rate is higher in sources with matching stellar objects.

\item
The correlation between thermal and non-thermal velocity dispersion in NH$_3$\,(1,1) indicates the dominance of non-thermal pressure and supersonic non-thermal motions in the dense gas traced by NH$_3$. In contrast to high-mass star-forming clumps and dense cores in star-forming regions, the relationship between non-thermal velocity dispersion and kinetic temperatures in PGCCs is notably weaker, with lower values observed for both parameters relative to other samples under our examination.

\item
The comparison of the line-center velocities of NH$_3$ with those from $^{13}$CO and C$^{18}$O reveals small discrepancies (0.17$\pm$0.33\,km\,s$^{-1}$, 0.12$\pm$0.18\,km\,s$^{-1}$). The widths of NH$_3$, $^{13}$CO, and C$^{18}$O in our sample were almost undistinguishable. These are consistent with the idea that these PGCCs are quiescent, as the majority of them appear to be transitioning from clouds to dense clumps.

\item
The ammonia column densities range between 0.36 to 6.07 $\times$ 10$^{15}$ cm$^{-2}$. The mean value is approximately 2.04 $\times$ 10$^{15}$ cm$^{-2}$, and the fractional abundances of ammonia range from 0.3 to 9.7 $\times$ 10$^{-7}$, with an average of 2.7 $\times$ 10$^{-7}$. Our observed fractional abundances of NH$_3$ are consistently one order of magnitude larger in PGCCs than in Massive Star-forming (MSF) regions and Infrared Dark Clouds (IRDCs)
This significant difference in NH$_3$ fractional abundance suggests that PGCCs are in an earlier phase of cloud evolution compared to MSF regions and IRDCs. The elevated NH$_3$ fractional abundance in PGCCs relative to other star-forming regions is likely attributable to the earlier phase of cloud evolution of PGCCs compared to Massive Star-forming (MSF) regions or even IRDCs.

\item
The kinetic temperatures of the cores range from 8.9 to 20.7\,K, with an average of 12.3 $\pm$ 2.9\,K. Similar temperature ranges between $T_{\rm kin}$ and $T_{\rm dust}$ indicate that the gas and dust are well coupled.

\item
The cumulative distribution shapes of line widths in the Planck cold cores closely resemble those of the dense cores found in regions Cepheus, and Orion L1630 and L1641, but with slightly higher values compared to Ophiuchus. However, the higher line width values in the Planck cold cores, when compared to these dense cores in Ophiuchus, suggest that they might be in more advanced and warmer stages of evolution. Nevertheless, it is worth noting that line width values remain small across the various samples under examination. This observation highlights the unique characteristics of the Planck cold cores in the context of their evolutionary stages.

\end{enumerate}

\begin{acknowledgements}
This work was mainly funded by the National Key R\&D Program of China under grant No. 2022YFA1603103. It was also partially funded by the Regional Collaborative Innovation Project of Xinjiang Uyghur Autonomous Region grant 2022E01050, the Tianshan Talent Program of Xinjiang Uygur Autonomous Region under grant No. 2022TSYCLJ0005, the Natural Science Foundation of Xinjiang Uygur Autonomous Region under grant No. 2022D01E06, the Chinese Academy of Sciences (CAS) “Light of West China” Program under grant Nos. xbzg-zdsys-202212, 2020-XBQNXZ-017, and 2021-XBQNXZ-028, the National Key R\&D Program of China with grant 2023YFA1608002, the National Natural Science Foundation of China (NSFC) under grants Nos. 12173075, 12373029, and 12103082, the Xinjiang Key Laboratory of Radio Astrophysics under grant No. 2023D04033, the Youth Innovation Promotion Association CAS. C.H. has been funded by the Chinese Academy of Sciences President’s International Fellowship Initiative by grant No. 2023VMA0031. Moreover, this work is sponsored (in part) by the Science Committee of the Ministry of Science and Higher Education of the Republic of Kazakhstan grant No. AP13067768. This work is based on observations made with the Nanshan 26-m radio telescope, the Nanshan 26-m Radio Telescope is partly supported by the Operation, Maintenance and Upgrading Fund for Astronomical Telescopes and Facility Instruments, budgeted from the Ministry of Finance of China (MOF) and administrated by the Chinese Academy of Sciences (CAS), and the Urumqi Nanshan Astronomy and Deep Space Exploration Observation and Research Station of Xinjiang.

\end{acknowledgements}

\onecolumn
\begin{appendix} 
\section{ECC Clump Catalogue}
\label{appendixA}
\begin{table}[h!]
\caption{Basic information of the observed sources} 
\label{table1}
\centering
\tiny
\begin{tabular}  {|p{0.35cm} p{1.71cm} p{0.7cm} p{0.74cm} p{0.6cm} |p{0.35cm} p{1.71cm} p{0.7cm} p{0.74cm} p{0.6cm}| p{0.35cm} p{1.71cm} p{0.7cm} p{0.74cm} p{0.6cm}|}
\hline 
 Num & Name & Glon & Glat & Dist & Num & Name & Glon & Glat & Dist & Num & Name & Glon & Glat & Dist \\ 
 & & $^{\circ}$ & $^{\circ}$ & kpc &  & & $^{\circ}$ & $^{\circ}$ & kpc &  & & $^{\circ}$ & $^{\circ}$ & kpc \\
\hline
1    &  G005.03+19.07  &  5.03    &   19.08  &  1.50  &    71   &  G052.99+03.07  &  53.00   &   3.08   &        &     141  &  G098.50-03.24  &  98.50   &   -3.25  &        \\
2    &  G005.29+14.47  &  5.30    &   14.48  &        &    72   &  G054.03-02.38  &  54.03   &   -2.39  &        &     142  &  G098.52+13.88  &  98.53   &   13.88  &  1.22  \\
3    &  G005.29+11.07  &  5.30    &   11.07  &  1.44  &    73   &  G056.84+04.81  &  56.84   &   4.82   &        &     143  &  G098.89+14.01  &  98.90   &   14.02  &  1.17  \\
4    &  G005.31+10.78  &  5.32    &   10.79  &  0.12  &    74   &  G057.08+04.46  &  57.08   &   4.46   &        &     144  &  G098.96+13.68  &  98.96   &   13.69  &  1.17  \\
5    &  G005.69+36.84  &  5.69    &   36.84  &  0.21  &    75   &  G057.10+03.65  &  57.11   &   3.66   &  0.77  &     145  &  G099.11+13.97  &  99.12   &   13.98  &        \\
6    &  G005.80+19.92  &  5.80    &   19.93  &  1.05  &    76   &  G057.17+03.41  &  57.17   &   3.42   &        &     146  &  G099.97+14.76  &  99.98   &   14.77  &        \\
\hline
\end{tabular}
\end{table}

\begin{table}[th!]
\caption{Continued} 
\label{table1}
\centering
\tiny
\begin{tabular}  {|p{0.35cm} p{1.71cm} p{0.7cm} p{0.74cm} p{0.6cm} |p{0.35cm} p{1.71cm} p{0.7cm} p{0.74cm} p{0.6cm}| p{0.35cm} p{1.71cm} p{0.7cm} p{0.74cm} p{0.6cm}|}
\hline \hline 
 Num & Name & Glon & Glat & Dist & Num & Name & Glon & Glat & Dist & Num & Name & Glon & Glat & Dist \\ 
 & & $^{\circ}$ & $^{\circ}$ & kpc &  & & $^{\circ}$ & $^{\circ}$ & kpc &  & & $^{\circ}$ & $^{\circ}$ & kpc \\
\hline
211  &  G117.11+12.42  &  117.11  &   12.43  &  0.11  &    286  &  G144.44+08.55  &  144.45  &   8.56   &  -0.66 &     361  &  G159.41-34.36  &  159.41  &   -34.36 &  1.20  \\
212  &  G118.08-09.68  &  118.08  &   -9.69  &        &    287  &  G144.49+08.70  &  144.49  &   8.71   &        &     362  &  G159.52+03.26  &  159.52  &   3.27   &  3.50  \\
213  &  G118.14+08.83  &  118.15  &   8.84   &        &    288  &  G144.66+00.16  &  144.67  &   0.17   &  0.85  &     363  &  G159.58-32.84  &  159.59  &   -32.84 &  0.25  \\
214  &  G118.25-52.70  &  118.25  &   -52.71 &  0.26  &    289  &  G144.66+06.11  &  144.67  &   6.11   &  0.53  &     364  &  G159.65-19.68  &  159.65  &   -19.69 &  0.29  \\
215  &  G118.34+08.66  &  118.34  &   8.67   &        &    290  &  G144.84+00.76  &  144.84  &   0.76   &        &     365  &  G159.65+11.39  &  159.65  &   11.40  &  0.17  \\
216  &  G118.36+21.74  &  118.37  &   21.74  &  0.38  &    291  &  G145.08-39.30  &  145.09  &   -39.31 &  0.28  &     366  &  G159.67-05.71  &  159.68  &   -5.72  &  0.14  \\
\hline
\end{tabular}
\end{table}

\begin{table}[th!]
\caption{Continued} 
\label{table1}
\centering
\tiny
\begin{tabular}  {|p{0.35cm} p{1.71cm} p{0.7cm} p{0.74cm} p{0.6cm} |p{0.35cm} p{1.71cm} p{0.7cm} p{0.74cm} p{0.6cm}| p{0.35cm} p{1.71cm} p{0.7cm} p{0.74cm} p{0.6cm}|}
\hline \hline 
 Num & Name & Glon & Glat & Dist & Num & Name & Glon & Glat & Dist & Num & Name & Glon & Glat & Dist \\ 
 & & $^{\circ}$ & $^{\circ}$ & kpc &  & & $^{\circ}$ & $^{\circ}$ & kpc &  & & $^{\circ}$ & $^{\circ}$ & kpc \\
\hline
436  &  G171.14-17.58  &  171.15  &   -17.58 &  0.79  &    511  &  G177.86+01.04  &  177.87  &   1.04   &  1.71  &     586  &  G206.54-14.45  &  206.54  &   -14.46 &  0.42  \\ 
437  &  G171.32+04.40  &  171.32  &   4.41   &        &    512  &  G177.89-20.16  &  177.89  &   -20.17 &  0.36  &     587  &  G206.60-14.40  &  206.61  &   -14.40 &  0.62  \\ 
438  &  G171.34-10.67  &  171.34  &   -10.67 &  0.23  &    513  &  G177.97-09.72  &  177.98  &   -9.73  &  0.60  &     588  &  G206.87-04.36  &  206.87  &   -4.37  &  0.42  \\ 
439  &  G171.34+02.59  &  171.34  &   2.59   &  1.78  &    514  &  G178.28-00.61  &  178.29  &   -0.62  &  0.94  &     589  &  G207.35-19.82  &  207.36  &   -19.83 &  0.42  \\ 
440  &  G171.43-17.36  &  171.43  &   -17.37 &  0.23  &    515  &  G178.48-06.76  &  178.48  &   -6.77  &  0.23  &     590  &  G208.56+02.36  &  208.56  &   2.37   &  1.14  \\ 
441  &  G171.49-14.90  &  171.50  &   -14.90 &  1.47  &    516  &  G178.72-07.01  &  178.73  &   -7.01  &  0.23  &     591  &  G209.20+02.12  &  209.20  &   2.13   &  1.13  \\
\hline
\end{tabular}
\end{table}

\begin{table}[th!]
\caption{Continued} 
\label{table1}
\centering
\tiny
\begin{tabular}  {|p{0.35cm} p{1.71cm} p{0.7cm} p{0.74cm} p{0.6cm} |p{0.35cm} p{1.71cm} p{0.7cm} p{0.74cm} p{0.6cm}| p{0.35cm} p{1.71cm} p{0.7cm} p{0.74cm} p{0.6cm}|}
\hline \hline 
 Num & Name & Glon & Glat & Dist & Num & Name & Glon & Glat & Dist & Num & Name & Glon & Glat & Dist \\ 
 & & $^{\circ}$ & $^{\circ}$ & kpc &  & & $^{\circ}$ & $^{\circ}$ & kpc &  & & $^{\circ}$ & $^{\circ}$ & kpc \\
\hline
661  &  G236.68-05.21  &  236.69  &   -5.21  &        &    665  &  G238.71-01.47  &  238.71  &   -1.47  &  1.13  &     669  &  G353.54+15.73  &  353.54  &   15.73  &  0.12   \\
662  &  G237.30-04.76  &  237.30  &   -4.76  &        &    666  &  G239.58-04.61  &  239.59  &   -4.61  &        &     670  &  G354.17+15.32  &  354.18  &   15.33  &  0.12   \\
663  &  G238.44-04.16  &  238.45  &   -4.16  &        &    667  &  G240.99-01.21  &  241.00  &   -1.21  &  6.15  &     671  &  G354.19+16.27  &  354.20  &   16.28  &  0.12   \\
664  &  G238.51-03.88  &  238.51  &   -3.88  &  1.00  &    668  &  G353.49+15.57  &  353.50  &   15.58  &  0.12  &     672  &  G354.46+16.15  &  354.46  &   16.16  &  0.12   \\

\hline
\end{tabular}
\end{table}

\setcounter{table}{1}
\begin{table}
\tiny
\centering
\caption{Observed parameters of the NH$_3$\,(1,1) and NH$_3$\,(2,2 ) emission lines in the NH$_3$ detected sources.}
\label{NH3_line}
\begin{tabular}{lccccccc}
\hline
Name  & $V_{\rm LSR}(1,1)$ & $\Delta v(1,1)$ & $\tau_{\rm}(1,1)$ & $T_{\rm MB}(1,1)$ &$V_{\rm LSR}(2,2)$ & $\Delta v(2,2)$ & $T_{\rm MB}(2,2)$ \\
 &  km\,s$^{-1}$ & km\,s$^{-1}$ & &K  &  km\,s$^{-1}$ & km\,s$^{-1}$ & K \\
\hline
G005.03+19.07  &   3.88(0.03)   &  0.76(0.06)  &              &   0.07(0.01)  &                &              &               \\
G005.29+11.07  &   4.04(0.03)   &  0.83(0.08)  &  0.64(0.47)  &   0.12(0.01)  &                &              &               \\
G005.69+36.84  &   1.76(0.04)   &  0.48(0.08)  &              &   0.03(0.01)  &                &              &               \\
G005.80+19.92  &   2.79(0.04)   &  0.65(0.09)  &              &   0.27(0.04)  &                &              &               \\
G006.04+36.74  &   2.37(0.02)   &  0.68(0.04)  &  4.16(0.30)  &   1.11(0.09)  &   2.30(0.03)   &  0.32(0.06)  &   0.20(0.03)  \\
G006.08+20.26  &   3.96(0.05)   &  0.92(0.14)  &              &   0.21(0.04)  &                &              &               \\
\hline
\end{tabular}
\end{table}

\setcounter{table}{1}
\begin{table}
\tiny
\centering
\caption{Continued}
\label{NH3_line}
\begin{tabular}{lccccccc}
\hline
Name  & $V_{\rm LSR}(1,1)$ & $\Delta v(1,1)$ & $\tau_{\rm}(1,1)$ & $T_{\rm MB}(1,1)$ &$V_{\rm LSR}(2,2)$ & $\Delta v(2,2)$ & $T_{\rm MB}(2,2)$ \\
 &  km\,s$^{-1}$ & km\,s$^{-1}$ & &K  &  km\,s$^{-1}$ & km\,s$^{-1}$ & K \\
\hline
G097.20+09.87  &   -2.59(0.04)  &  1.02(0.08)  &  1.92(0.47)  &   0.25(0.03)  &                &              &               \\
G097.38+09.95  &   -2.38(0.03)  &  0.75(0.06)  &  3.20(0.67)  &   0.12(0.01)  &                &              &               \\
G098.52+13.88  &   -3.21(0.02)  &  0.69(0.05)  &  0.88(0.50)  &   0.13(0.01)  &                &              &               \\
G098.89+14.01  &   -2.85(0.02)  &  0.71(0.04)  &  2.37(0.34)  &   0.22(0.02)  &   -2.73(0.04)  &  0.28(0.10)  &   0.04(0.01)  \\
G098.96+13.68  &   -2.91(0.03)  &  0.62(0.07)  &  1.60(0.85)  &   0.05(0.01)  &                &              &               \\
G102.34+15.96  &   2.44(0.02)   &  0.76(0.04)  &  2.98(0.15)  &   0.72(0.05)  &   2.39(0.05)   &  0.66(0.10)  &   0.08(0.02)  \\
\hline
\end{tabular}
\end{table}

\setcounter{table}{1}
\begin{table}
\tiny
\centering
\caption{Continued}
\label{NH3_line}
\begin{tabular}{lccccccc}
\hline
Name  & $V_{\rm LSR}(1,1)$ & $\Delta v(1,1)$ & $\tau_{\rm}(1,1)$ & $T_{\rm MB}(1,1)$ &$V_{\rm LSR}(2,2)$ & $\Delta v(2,2)$ & $T_{\rm MB}(2,2)$ \\
 &  km\,s$^{-1}$ & km\,s$^{-1}$ & &K  &  km\,s$^{-1}$ & km\,s$^{-1}$ & K \\
\hline
G160.53-19.72  &   -3.84(0.02)  &  0.99(0.05)  &  0.58(0.23)  &   0.38(0.03)  &                &              &               \\
G160.53-09.84  &   3.50(0.02)   &  0.87(0.05)  &              &   0.11(0.01)  &                &              &               \\
G161.21-08.72  &   -3.86(0.02)  &  0.62(0.05)  &  1.68(0.65)  &   0.22(0.02)  &                &              &               \\
G161.43-35.59  &   -5.87(0.05)  &  0.87(0.10)  &              &   0.12(0.02)  &                &              &               \\
G161.60-08.34  &   -3.26(0.03)  &  0.68(0.08)  &              &   0.12(0.02)  &                &              &               \\
G161.85-08.66  &   -2.50(0.03)  &  0.70(0.06)  &  2.54(0.71)  &   0.22(0.03)  &                &              &               \\
\hline
\end{tabular}
\end{table}

\setcounter{table}{1}
\begin{table}
\tiny
\centering
\caption{Continued}
\label{NH3_line}
\begin{tabular}{lccccccc}
\hline
Name  & $V_{\rm LSR}(1,1)$ & $\Delta v(1,1)$ & $\tau_{\rm}(1,1)$ & $T_{\rm MB}(1,1)$ &$V_{\rm LSR}(2,2)$ & $\Delta v(2,2)$ & $T_{\rm MB}(2,2)$ \\
 &  km\,s$^{-1}$ & km\,s$^{-1}$ & &K  &  km\,s$^{-1}$ & km\,s$^{-1}$ & K \\
\hline
G192.94-03.73  &   4.00(0.05)   &  1.36(0.12)  &              &   0.05(0.01)  &                &              &               \\
G194.80-03.41  &   12.14(0.04)  &  1.24(0.10)  &  1.25(0.35)  &   0.11(0.01)  &                &              &               \\
G198.03-15.24  &   -0.13(0.03)  &  0.94(0.07)  &  0.40(0.43)  &   0.07(0.01)  &                &              &               \\
G198.56-09.10  &   11.25(0.02)  &  0.72(0.05)  &  1.39(0.44)  &   0.41(0.04)  &                &              &               \\
G199.88+00.93  &   5.70(0.03)   &  0.98(0.07)  &  1.42(0.35)  &   0.33(0.04)  &                &              &               \\
G201.13+00.31  &   4.75(0.04)   &  0.90(0.08)  &  1.24(0.60)  &   0.15(0.02)  &                &              &               \\
\hline
\end{tabular}
\end{table}

\setcounter{table}{2}
\begin{table}[th!]
\tiny
\centering
\caption{Calculated parameters of the NH$_3$\,(1,1) and (2,2 ) emission lines detected in the 73 cores.}
\label{NH_para}
\begin{tabular}{cccccccccccccccc}
\hline \hline
Name &   T$_{\rm ex}$ & T$_{\rm rot}$ & T$_{\rm kin}$ & $N_{\rm NH_{3}}$ & $\sigma_{\rm NT}$ & $\sigma_{\rm TH}$ & c$_{\rm s}$ & $\mathcal{M}$ &$R_{\rm P}$ &   $\chi$  \\
 & K & K & K & 10$^{15}$cm$^{-2}$ & km\,s$^{-1}$ & km\,s$^{-1}$  &  km\,s$^{-1}$  & &  & 10$^{-7}$   \\
\hline
G006.04+36.74   &   3.8(1.1) &	 9.6(0.1)  &	 10.0(0.1) &	3.6(0.2) &	0.40(0.01) &	0.07(0.01) &	0.19(0.01) &	2.2(0.2) &	0.03(0.01) &      ...  \\
G006.32+20.44   &   3.9(1.2) &	 12.3(0.1) &	 13.4(0.1) &	0.7(0.1) &	0.41(0.01) &	0.08(0.01) &	0.22(0.01) &	1.9(0.1) &	0.04(0.01) & 0.6(0.1)  \\
G006.70+20.66   &   3.9(0.8) &	 13.0(0.4) &	 14.2(0.5) &	0.4(0.1) &	0.50(0.01) &	0.08(0.01) &	0.22(0.01) &	2.3(0.1) &	0.03(0.01) & 0.3(0.1)  \\
G006.98+20.72   &   2.9(0.1) &	 17.2(0.4) &	 20.2(0.6) &	0.9(0.3) &	0.41(0.01) &	0.10(0.01) &	0.27(0.01) &	1.6(0.1) &	0.06(0.01) & 1.4(0.3)  \\
G007.80+21.10   &   2.8(0.1) &	 14.4(0.3) &	 16.1(0.4) &	1.4(0.5) &	0.53(0.02) &	0.09(0.01) &	0.24(0.01) &	2.3(0.7) &	0.03(0.01) & 2.5(0.6)  \\
G008.67+22.14   &   3.3(0.6) &	 9.4(0.1)  &	 9.8(0.1)  &	3.3(0.1) &	0.39(0.01) &	0.07(0.01) &	0.18(0.01) &	2.1(0.2) &	0.03(0.01) & 4.6(0.1)  \\
\hline
\end{tabular}
\end{table}

\setcounter{table}{3}
\begin{table}[th!]
\tiny\tiny
\centering
\caption{Kolmogorov-Smirnov test results for line widths, kinematic temperatures, columnn densities and abundances distributions in different star formation samples.}
\label{ks-test}
\begin{tabular}{lcccc}
\hline \hline
Distributions &  Test Parameter  & KS Statistic & P-Value    \\
\hline
Planck cold cores and  BGPS                         &  $T_{\rm kin}$              & 0.44  & 3.23 $\times$ $10^{-08}$ \\
Planck cold cores and  ATLASGAL                     &  $T_{\rm kin}$              & 0.75  & 3.65 $\times$ $10^{-34}$ \\
Planck cold cores and  UC H{\sc ii} Cand            &  $T_{\rm kin}$              & 0.85  & 1.40 $\times$ $10^{-18}$ \\
Planck cold cores and  High contrast IRDCs          &  $T_{\rm kin}$              & 0.50  & 4.58 $\times$ $10^{-17}$ \\
Planck cold cores and  High IR extinction clouds    &  $T_{\rm kin}$              & 0.78  & 1.54 $\times$ $10^{-15}$ \\
Planck cold cores and  EGOs                         &  $T_{\rm kin}$              & 0.67  & 9.29 $\times$ $10^{-05}$ \\
Planck cold cores and  Ophichus                     &  $T_{\rm kin}$              & 0.46  & 1.34 $\times$ $10^{-02}$ \\
\hline
\end{tabular}
\end{table}

\setcounter{table}{4}
\begin{table}[th!]
\tiny\tiny
\centering
\caption{Spearman correlation of kinetic temperature versus non-thermal velocity dispersion in different star formation samples.}
\label{Spearman-test}
\begin{tabular}{lcc}
\hline \hline
Dataset &  Spearman Correlation & P-value    \\
\hline
ATLASGAL                      &   0.53                  &  1.48 $\times$ $10^{-53}$  \\
BGPS                          &   0.52                  &  1.20 $\times$ $10^{-36}$  \\
High contrast IRDCs           &   0.07                  &  0.45                     \\
Ophiuchus                     &   0.49                  &  0.03                     \\
Cepheus \& Orion L1630/L1641  &   0.49                  &  2.01 $\times$ $10^{-4}$   \\
Perseus                       &   0.2                   &  0.16                     \\
Planck cold cores             &   0.33                  &  4.00 $\times$ $10^{-3}$   \\
\hline
\end{tabular}
\end{table}

\twocolumn
\section{Samples of NH$_3$ spectra}
\label{appendB-spec}

\begin{figure}[h]
\vspace*{0.2mm}
\centering
\includegraphics[width=0.45\textwidth]{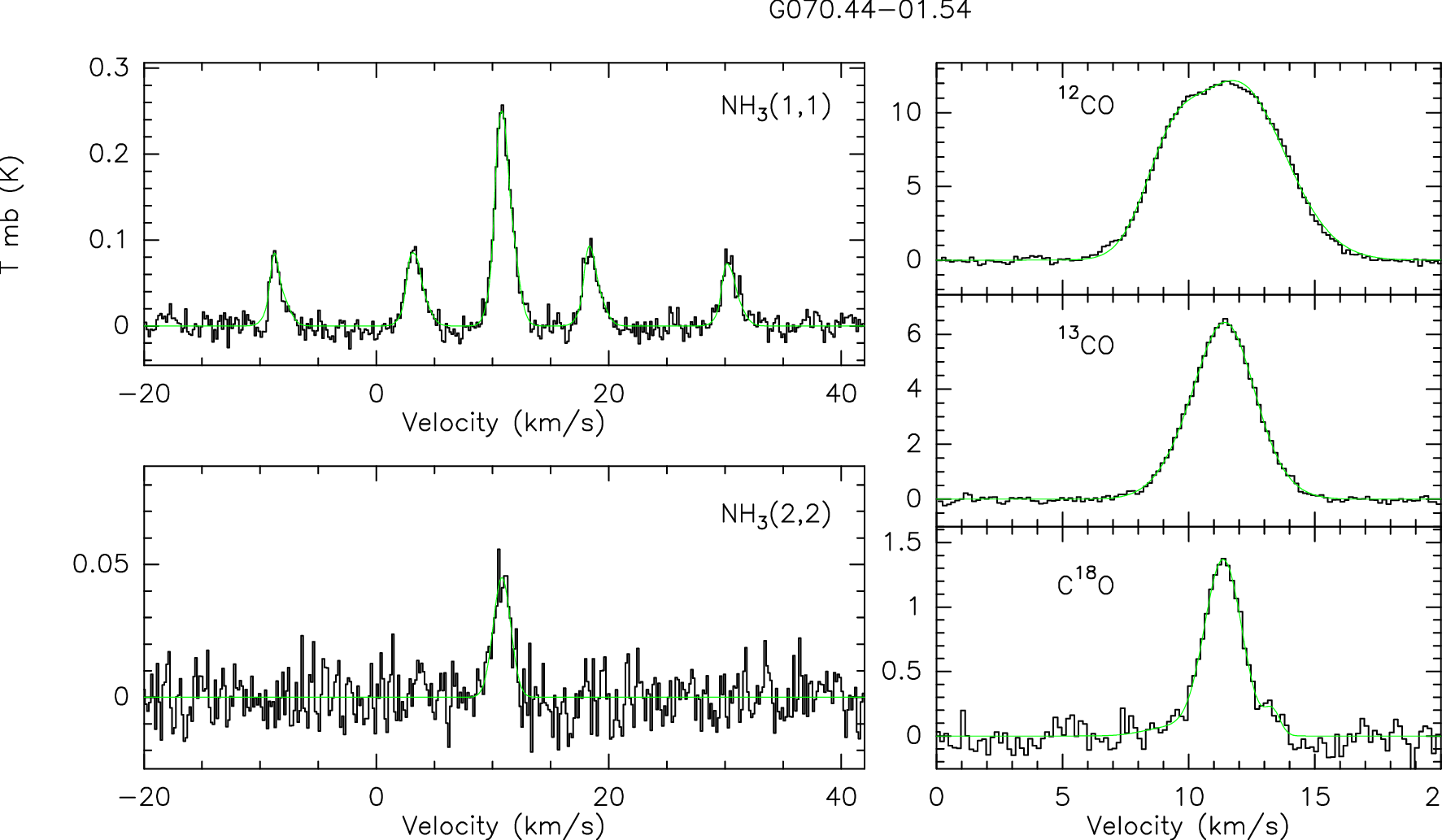}

\includegraphics[width=0.45\textwidth]{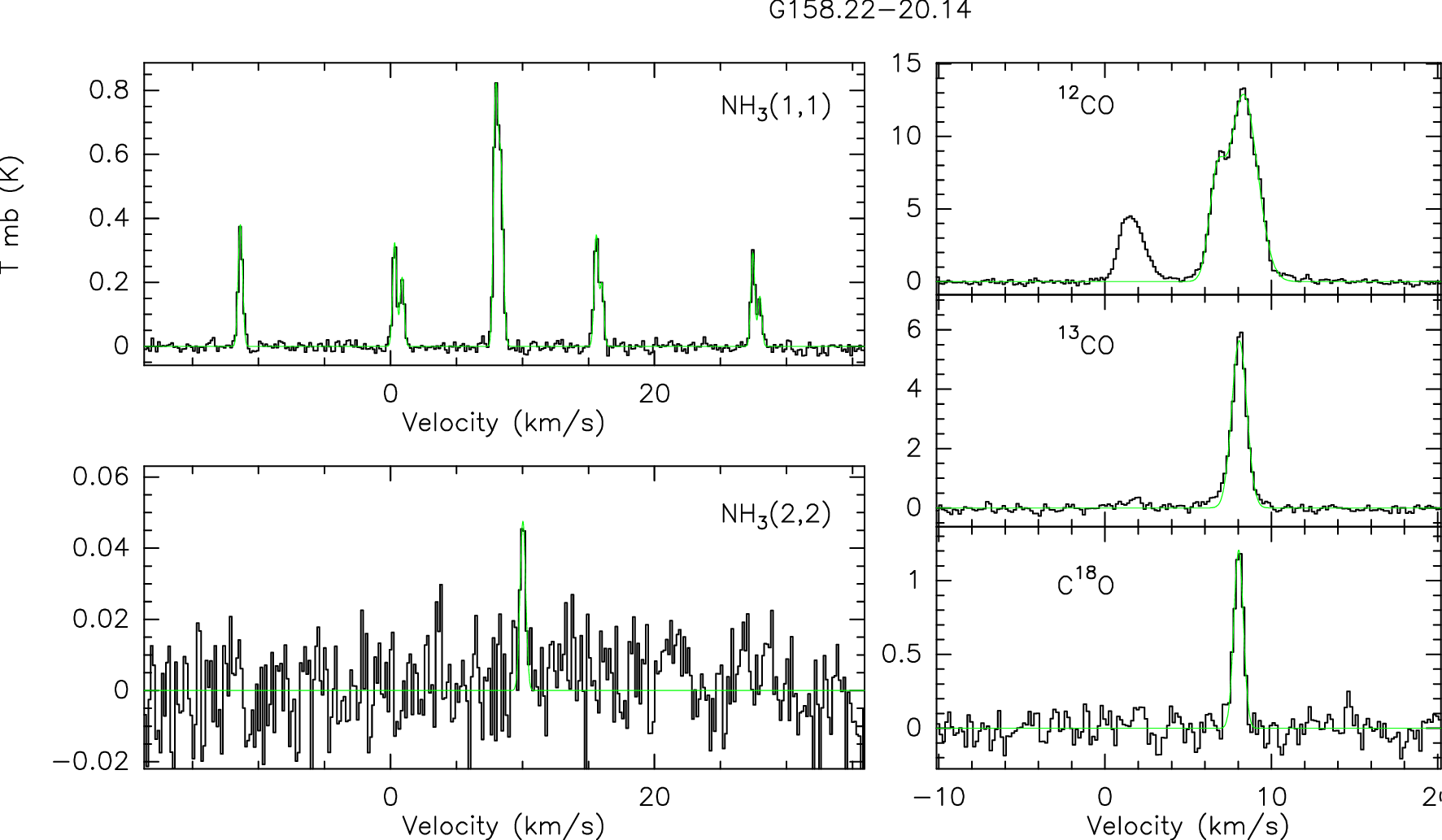}

\includegraphics[width=0.45\textwidth]{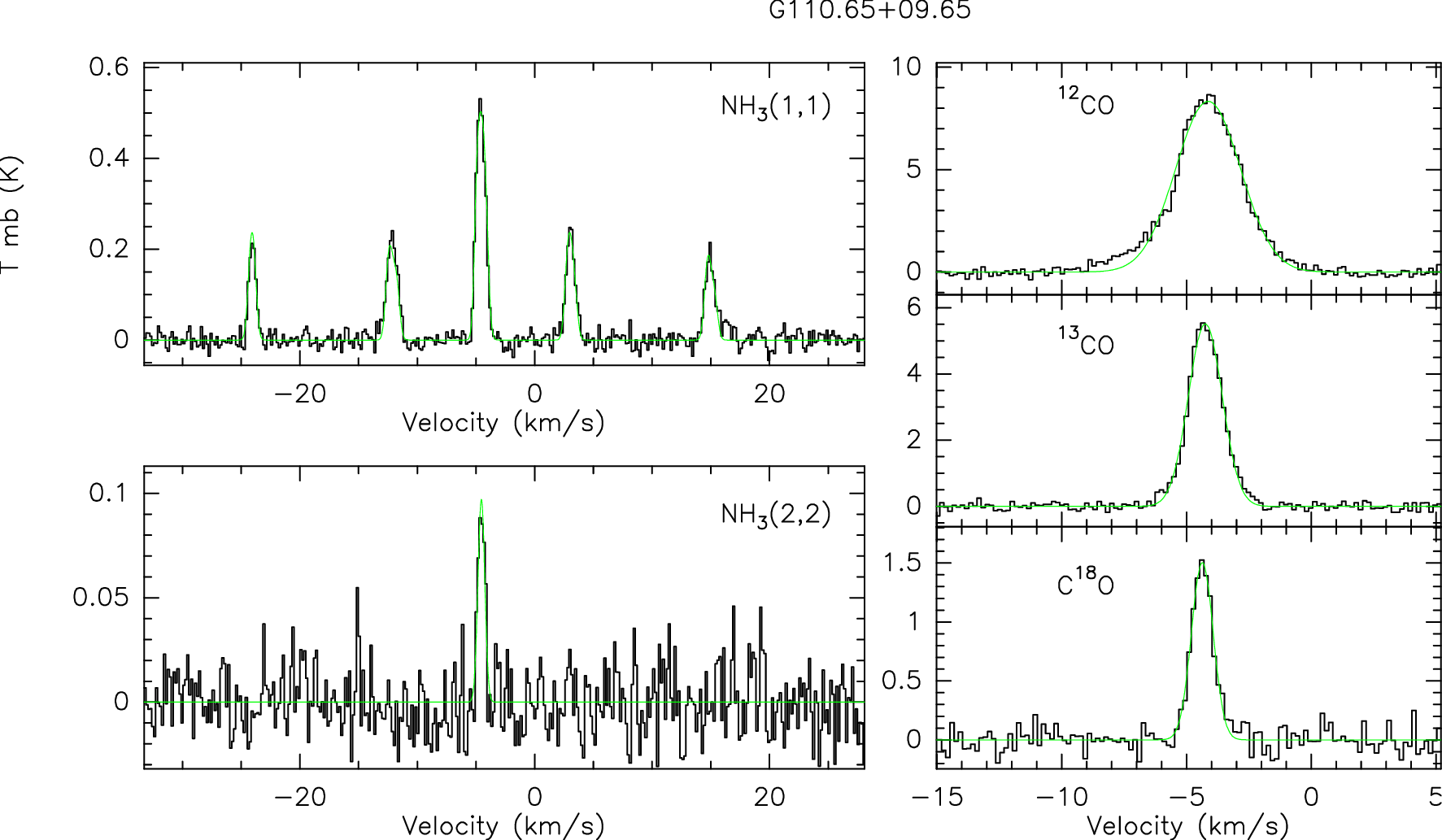}

\includegraphics[width=0.45\textwidth]{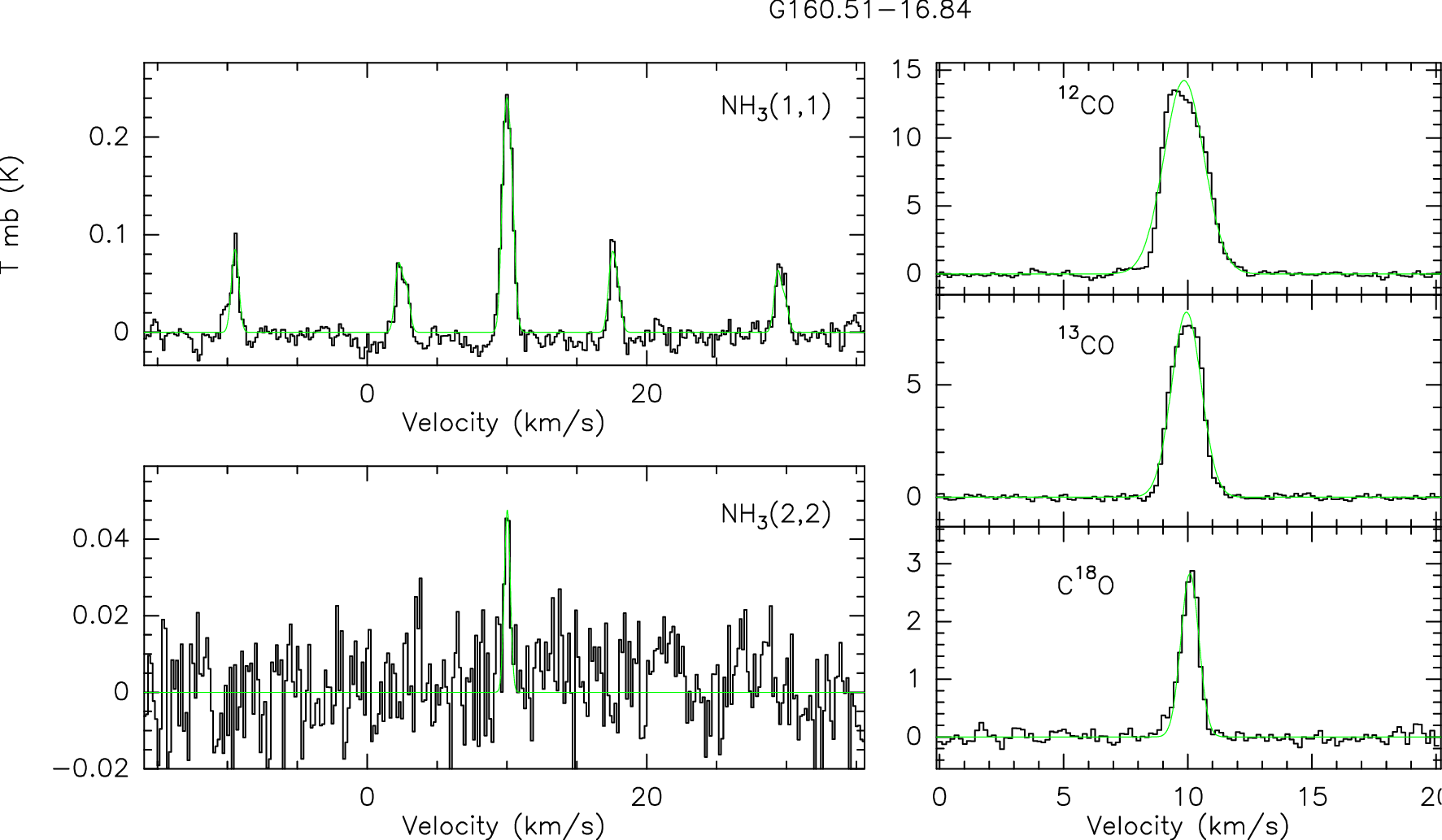}
\caption[]{Examples for typical ammonia spectra in different S/N ratios. Spectra include NH$_3$(1,1) and (2,2), $^{12}$CO, $^{13}$CO and C$^{18}$O (1-0) transitions towards cores, namely G070.44-01.54, G110.65+09.65, G158.22-20.14 and G160.51-16.84. The green color indicates the NH$_3$(1,1) fitting and Gauss fitting of NH$_3$(2,2), $^{12}$CO, $^{13}$CO and C$^{18}$O profile.}
\label{b.1}
\end{figure}

\section{Derived CO parameters}
\label{appendB}

Our calculations assume that the cores are in a state of local thermodynamical equilibrium (LTE) and the $^{13}$CO emission is optically thin. Adopting a beam-filling factor of unity, the excitation temperature for CO is derived from the following equation \citep{2010ApJ...721..686P,2015ApJ...805...58K}
\begin{equation}\label{}
\begin{split}
T_{\rm ex (CO)} =5.33 \left\{ \rm ln \left[1+ \frac{5.33}{\left(T_{\rm mb,^{12}\rm CO}+0.818\right)}\right] \right\}^{-1}\, \rm K,
\end{split}
\end{equation}  
where $T_{\rm mb,^{12}\rm CO}$ is the peak intensity of $^{12}$CO (J = 1–0) in units of K. We then obtain the optical depths, $\tau$, and the column densities, $N_{\rm ^{13}CO}$, $N_{\rm C^{18}O}$ for the $^{13}$CO and C$^{18}$O molecules using the following equations \citep{1994ApJ...429..694L,1998ApJS..117..387K}.
\begin{equation}\label{}
	\begin{aligned}
		\tau_{\rm ^{13}CO} & =-\ln \left\{1 - \frac{T_{\rm mb,^{13}CO}}{5.29([\exp(5.29 / T_{\rm ex (CO)}-1]^{-1}-0.164)}\right\}\,,
	\end{aligned}
\end{equation}

\begin{equation}\label{}
	\begin{aligned}
		\tau_{\rm C^{18}O} & =-\ln \left\{1 - \frac{T_{\rm mb,C^{18}O}}{5.27([\exp(5.27 / T_{\rm ex (CO)}-1]^{-1}-0.166)}\right\}\,,
	\end{aligned}
\end{equation}

\begin{equation}\label{}
	\begin{aligned}
		N_{\rm ^{13}CO} & =2.42\times 10^{14} \frac{\tau_{\rm ^{13}CO} \Delta v_{\rm ^{13}CO} T_{\rm ex (CO)}}{1-\exp(-5.29/T_{\rm ex (CO)})}\, \rm cm^{-2},
	\end{aligned}
\end{equation}
and 
\begin{equation}\label{}
	\begin{aligned}
		N_{\rm C ^{18}O} & =2.42\times 10^{14} \frac{\tau_{\rm C^{18}O} \Delta v_{\rm C^{18}O} T_{\rm ex (CO)}}{1-\exp(-5.27/T_{\rm ex (CO)})}\, \rm cm^{-2},
	\end{aligned}
\end{equation}

where $T_{\rm mb,^{13}\rm CO}$ and $T_{\rm mb,\rm C^{18}O}$ are peak intensities in K, while $\Delta v_{\rm ^{13}CO}$ and $\Delta V_{\rm C^{18}O}$ are the FWHM inewidths in\,km\,s$^{-1}$. In the calculation, the excitation temperatures of $^{13}$CO have been assigned to be equal to those of $^{12}$CO. This would be reasonable when considering the fact that $^{12}$CO, $^{13}$CO, and $C^{18}$O are well coupled in ECCs \citep{2012ApJ...756...76W}. Molecular hydrogen column densities were calculated by the fractional abundances of [H$_{2}$]/[$^{13}$CO] = 89 × 10$^{4}$ \citep{1980ApJ...237....9M} and [H$_{2}$]/[C$^{18}$O] = 7 × 10$^{6}$ \citep{1982ApJ...262..590F} in the solar neighborhood. 

We find that the hydrogen column density ($N\rm _{H_{2}}$) values derived from $^{13}$CO and C$^{18}$O are quite close to each other, suggesting that both $^{13}$CO and C$^{18}$O are optically thin and the optical depth effect can be ignored in our calculation \citep{2013ApJS..209...37M}. Therefore we use $^{13}$CO to calculate $N\rm _{H_{2}}$, as listed in Table\,\ref{CO_cal} of Appendix\,\ref{appendB}.

The excitation temperatures of CO $J$ = 1-0 ($T_{\rm ex (CO)}$) for our cores range from 4.4 to 19.7\,K. The mean value of $T_{\rm ex (CO)}$) is 10.5\,K and it is smaller than the value in \cite{2012ApJ...756...76W} and the average dust temperature (13\,K) for C3POs \citep{2011A&A...536A...1P}. The H$\rm _{2}$ column densities $N\rm _{H_{2}}$($^{13}$CO) of our cores are derived from $N_{\rm ^{13}CO}$ covering the range of (0.07-2.88) $\times$ 10$^{22}$cm$^{-2}$ with a mean value of 7.4 $\times$ 10$^{21}$ cm$^{-2}$. Sources in our sample have similar $N\rm _{H_{2}}$ column densities to those in Galactic second quadrant samples, with a mean value of 8 $\times$ 10$^{21}$ cm$^{-2}$ \citep{2013ApJS..209...37M}.

\onecolumn
\centering
\setcounter{table}{0}
\tiny\tiny
\begin{table}   
\caption{Parameters derived from the $^{12}$CO, $^{13}$CO and C$^{18}$O (1-0) lines}
\label{CO_line}
\begin{tabular}{lccccccccccccccc}
\hline
Name &   $V_{\rm LSR}$(CO) & $\Delta v_{\rm CO}$ & $T_{\rm mb,^{12}\rm CO}$ & $V_{\rm LSR}(\rm^{13}CO)$ & $\Delta v_{\rm ^{13}CO}$ & $T_{\rm mb,^{13}\rm CO}$ & $V_{\rm LSR}(\rm C^{18}O)$ & $\Delta v_{\rm C^{18}O}$ & $T_{\rm mb,\rm C^{18}O}$\\
 & km\,s$^{-1}$& km\,s$^{-1}$&  K & km\,s$^{-1}$& km\,s$^{-1}$ & K & km\,s$^{-1}$ & km\,s$^{-1}$ & K   \\
\hline
G005.03+19.07	&  3.7(0.1)  & 	1.3(0.1) 	&  14.5(0.2) &  3.7(0.1)  &  1.0(0.1) &  8.1(0.1)  &   3.7(0.1) &  	0.7(0.1) 	&  1.6(0.2)  & \\
G005.29+11.07	&  4.3(0.1)  & 	0.9(0.1) 	&  16.2(0.3) &  4.2(0.1)  &  0.6(0.1) &  7.9(0.1)  &   4.2(0.1) &  	0.5(0.1) 	&  1.6(0.1)  & \\
G005.80+19.92	&  2.7(0.1)  & 	0.9(0.1) 	&  6.2(0.3)  &  2.8(0.1)  &  0.7(0.1) &  6.4(0.1)  &   2.9(0.1) &  	0.5(0.1) 	&  1.2(0.1)  & \\
G006.08+20.26	&  4.0(0.1)  & 	1.6(0.1) 	&  10.3(0.2) &  3.9(0.1)  &  1.0(0.1) &  8.6(0.1)  &   3.9(0.1) &  	0.6(0.1) 	&  2.0(0.2)  & \\
G006.32+20.44	&  4.3(0.1)  & 	1.7(0.1) 	&  10.9(0.2) &  4.3(0.1)  &  0.8(0.1) &  8.2(0.2)  &   4.2(0.1) &  	0.5(0.1) 	&  3.3(0.1)  & \\
G006.41+20.56	&  4.7(0.1)  & 	1.1(0.1) 	&  12.7(0.3) &  4.6(0.2)  &  0.7(0.2) &  7.2(0.2)  &   4.4(0.1) &  	0.8(0.1) 	&  2.1(0.1)  & \\
\hline
\end{tabular}
\end{table}

\setcounter{table}{0}
\tiny\tiny
\begin{table}   
\caption{Continued}
\begin{tabular}{lccccccccccccccc}
\hline
Name &   $V_{\rm LSR}$(CO) & $\Delta v_{\rm CO}$ & $T_{\rm mb,^{12}\rm CO}$ & $V_{\rm LSR}(\rm^{13}CO)$ & $\Delta v_{\rm ^{13}CO}$ & $T_{\rm mb,^{13}\rm CO}$ & $V_{\rm LSR}(\rm C^{18}O)$ & $\Delta v_{\rm C^{18}O}$ & $T_{\rm mb,\rm C^{18}O}$\\
 & km\,s$^{-1}$& km\,s$^{-1}$&  K & km\,s$^{-1}$& km\,s$^{-1}$ & K & km\,s$^{-1}$ & km\,s$^{-1}$ & K   \\
\hline
G117.11+12.42	&  3.6(0.1)  & 	0.8(0.1) 	&  7.5(0.2)  &  3.9(0.1)  &  1.0(0.1) &  5.5(0.1)  &   3.9(0.1) & 	0.6(0.1) 	&  2.0(0.1)  & \\
G120.16+03.09	&  -19.1(0.1)& 	2.4(0.1) 	&  7.3(0.2)  &  -18.9(0.2)&  1.3(0.2) &  4.0(0.1)  & -19.3(0.1) & 	1.3(0.1) 	&  0.8(0.1)  & \\
G120.67+02.66	&  -18.0(0.1)& 	3.4(0.1) 	&  13.0(0.6) &  -18.0(0.1)&  2.2(0.1) &  5.3(0.1)  & -18.1(0.1) & 	1.8(0.1) 	&  0.7(0.1)  & \\
G120.98+02.66	&  -16.9(0.1)& 	2.7(0.1) 	&  9.0(0.6)  &  -16.9(0.1)&  1.7(0.1) &  3.8(0.1)  & -17.1(0.1) & 	1.5(0.2) 	&  0.5(0.1)  & \\
G121.35+03.39	&  -6.1(0.1) & 	2.3(0.1) 	&  4.8(0.4)  &  -5.5(0.1) &  1.5(0.1) &  3.5(0.1)  &   -5.4(0.1)&  	0.9(0.1) 	&  1.2(0.1)  & \\
G123.66+24.89	&  -4.2(0.2) & 	1.0(0.2) 	&  5.5(0.2)  &  -4.3(0.1) &  1.1(0.1) &  3.4(0.1)  &   -4.3(0.1)&  	0.7(0.1) 	&  0.6(0.1)  & \\
\hline
\end{tabular}
\end{table}

\setcounter{table}{0}
\tiny\tiny
\begin{table}   
\caption{Continued}
\begin{tabular}{lccccccccccccccc}
\hline
Name &   $V_{\rm LSR}$(CO) & $\Delta v_{\rm CO}$ & $T_{\rm mb,^{12}\rm CO}$ & $V_{\rm LSR}(\rm^{13}CO)$ & $\Delta v_{\rm ^{13}CO}$ & $T_{\rm mb,^{13}\rm CO}$ & $V_{\rm LSR}(\rm C^{18}O)$ & $\Delta v_{\rm C^{18}O}$ & $T_{\rm mb,\rm C^{18}O}$\\
 & km\,s$^{-1}$& km\,s$^{-1}$&  K & km\,s$^{-1}$& km\,s$^{-1}$ & K & km\,s$^{-1}$ & km\,s$^{-1}$ & K   \\
\hline
G165.34-07.51b	&  0.0(0.2)  & 	1.6(0.2) 	&  4.4(0.1)  &  -0.2(0.1) &  1.0(0.1) &  2.7(0.3)  &   -0.5(0.1)&  	1.0(0.1) 	&  0.9(0.1)  & \\
G165.69-09.14	&  -0.4(0.1) & 	2.8(0.2) 	&  4.7(0.6)  &  -0.6(0.1) &  1.5(0.1) &  4.0(0.3)  &   -0.8(0.1)&  	0.6(0.1) 	&  1.2(0.3)  & \\
G168.00-15.69	&  7.7(0.1)  & 	1.0(0.1) 	&  10.7(0.6) &  7.7(0.2)  &  0.8(0.2) &  4.7(0.1)  &   7.7(0.1) & 	0.6(0.1) 	&  2.5(0.1)  & \\
G168.13-16.39	&  5.9(0.2)  & 	0.9(0.2) 	&  7.6(0.2)  &  6.1(0.1)  &  0.6(0.1) &  4.0(0.1)  &   6.5(0.1) & 	0.8(0.1) 	&  2.7(0.1)  & \\
G168.72-15.48	&  7.1(0.2)  & 	1.3(0.2) 	&  9.1(0.4)  &  7.2(0.1)  &  1.4(0.1) &  6.4(0.3)  &   7.3(0.1) & 	0.6(0.1) 	&  2.7(0.2)  & \\
G169.76-16.15a	&  5.5(0.1)  & 	1.2(0.1) 	&  7.1(0.2)  &  5.5(0.1)  &  1.0(0.1) &  4.7(0.1)  &   5.5(0.1) & 	0.5(0.1) 	&  2.1(0.1)  & \\
\hline
\end{tabular}
\end{table}

\setcounter{table}{0}
\tiny\tiny
\begin{table}   
\caption{Continued}
\begin{tabular}{lccccccccccccccc}
\hline
Name &   $V_{\rm LSR}$(CO) & $\Delta v_{\rm CO}$ & $T_{\rm mb,^{12}\rm CO}$ & $V_{\rm LSR}(\rm^{13}CO)$ & $\Delta v_{\rm ^{13}CO}$ & $T_{\rm mb,^{13}\rm CO}$ & $V_{\rm LSR}(\rm C^{18}O)$ & $\Delta v_{\rm C^{18}O}$ & $T_{\rm mb,\rm C^{18}O}$\\
 & km\,s$^{-1}$& km\,s$^{-1}$&  K & km\,s$^{-1}$& km\,s$^{-1}$ & K & km\,s$^{-1}$ & km\,s$^{-1}$ & K   \\
\hline
G201.59+00.55b	&  4.9(0.1)  & 	3.8(0.1) 	&  8.5(0.2)  &  5.7(0.2)  &  1.1(0.2) &  3.0(0.1)  &   5.6(0.1) & 	0.7(0.2) 	&  0.5(0.1)  & \\
G202.30-08.91	&  12.0(0.1) & 	2.1(0.1) 	&  8.9(0.2)  &  12.0(0.1) &  1.7(0.1) &  4.9(0.6)  &   11.9(0.1)&  	1.0(0.1) 	&  1.1(0.1)  & \\
G203.20-11.20	&  10.6(0.1) & 	2.2(0.1) 	&  8.0(0.2)  &  10.6(0.1) &  1.6(0.1) &  6.3(0.2)  &   10.5(0.1)&  	1.2(0.1) 	&  2.4(0.1)  & \\
G203.75-08.49	&  12.1(0.2) & 	1.4(0.2) 	&  6.0(0.3)  &  11.6(0.1) &  1.0(0.1) &  5.0(0.1)  &   11.6(0.1)&  	0.7(0.1) 	&  1.1(0.1)  & \\
G204.82-13.88	&  7.3(0.3)  & 	3.6(0.3) 	&  8.4(0.2)  &  7.9(0.2)  &  1.2(0.2) &  3.5(0.1)  &   8.4(0.1) & 	1.0(0.1) 	&  2.0(0.1)  & \\
G205.57-08.25	&  11.8(0.1) & 	1.7(0.2) 	&  4.5(0.4)  &  11.5(0.1) &  1.6(0.1) &  3.3(0.2)  &   11.3(0.1)&  	1.1(0.3) 	&  0.9(0.3)  & \\
\hline
\end{tabular}
\end{table}

\onecolumn
\begin{sidewaystable}[th!]

\setcounter{table}{1}
\caption{Derived Parameters of CO lines}
\label{CO_cal}
\tiny
\begin{tabular}  {|p{1.8cm} p{0.8cm} p{0.9cm} p{0.9cm} p{0.9cm} p{0.9cm} p{1cm} p{1.1cm} |p{1.8cm} p{0.8cm} p{0.9cm} p{0.9cm} p{0.9cm} p{0.9cm} p{1cm} p{1.1cm} |}
\hline \hline 
 Name &$T_{\rm ex(CO)}$ & $\tau_{\rm^{13}CO}$  & $\tau_{\rm C^{18}O}$ & $N_{\rm ^{13}CO}$ & $N_{\rm C^{18}O}$ & $N_{\rm H_{2}}$($^{13}$CO) & $N_{\rm H_{2}}$(C$^{18}$O) & Name &$T_{\rm ex(CO)}$ & $\tau_{\rm^{13}CO}$  & $\tau_{\rm C^{18}O}$ & $N_{\rm ^{13}CO}$ & $N_{\rm C^{18}O}$ & $N_{\rm H_{2}}$($^{13}$CO) & $N_{\rm H_{2}}$(C$^{18}$O)\\ 
 & K & & & 10$^{15}$cm$^{-2}$ & 10$^{15}$cm$^{-2}$ & 10$^{22}$cm$^{-2}$ & 10$^{22}$cm$^{-2}$ & & K & & & 10$^{15}$cm$^{-2}$ & 10$^{15}$cm$^{-2}$ & 10$^{22}$cm$^{-2}$ & 10$^{22}$cm$^{-2}$  \\
\hline

G005.03+19.07  &  17.9(0.2)  &  0.8(0.01)  &  0.1(0.01) & 14.0(0.5)  &   1.4(0.2) &  1.2(0.4) &  1.0(0.1) &            G110.52+19.17  &  7.1(0.2)   &  1.4(0.03)  &  0.6(0.02) & 2.3(0.3)   &   0.7(0.1) &  0.2(0.3) &  0.5(0.1) \\
G005.29+11.07  &  19.6(0.3)  &  0.7(0.01)  &  0.1(0.01) & 8.6(0.3)   &   1.0(0.1) &  0.8(0.3) &  0.7(0.1) &            G110.65+09.65  &  11.6(0.2)  &  1.1(0.01)  &  0.2(0.01) & 13.0(0.7)  &   1.6(0.2) &  1.2(0.6) &  1.1(0.2) \\
G005.80+19.92  &  9.4(0.3)   &             &  0.2(0.58) &            &   0.6(1.8) &           &  0.4(1.2) &            G111.66+20.20  &  9.2(0.3)   &  2.8(0.20)  &  0.3(0.31) & 18.0(2.6)  &   1.2(1.5) &  1.6(2.3) &  0.8(1.0) \\
G006.08+20.26  &  13.6(0.2)  &  1.8(0.01)  &  0.2(0.01) & 18.0(0.7)  &   1.4(0.2) &  1.6(0.6) &  1.0(0.1) &            G111.68+19.74  &  6.7(0.3)   &  0.7(0.01)  &  0.4(0.01) & 1.6(0.3)   &   0.6(0.1) &  0.1(0.3) &  0.4(0.1) \\
G006.32+20.44  &  14.2(0.2)  &  1.4(0.01)  &  0.4(0.01) & 13.0(0.5)  &   2.2(0.2) &  1.1(0.5) &  1.5(0.1) &            G111.77+20.26  &  9.1(0.2)   &  1.0(0.01)  &  0.2(0.01) & 4.2(0.4)   &   0.8(0.1) &  0.4(0.3) &  0.6(0.1) \\
\hline
\end{tabular}
\end{sidewaystable}

\onecolumn
\begin{sidewaystable}[th!]
\setcounter{table}{1}
\caption{continued}
\tiny
\begin{tabular}  {|p{1.8cm} p{0.8cm} p{0.9cm} p{0.9cm} p{0.9cm} p{0.9cm} p{1cm} p{1.1cm} |p{1.8cm} p{0.8cm} p{0.9cm} p{0.9cm} p{0.9cm} p{0.9cm} p{1cm} p{1.1cm}|}
\hline \hline 
 Name &$T_{\rm ex(CO)}$ & $\tau_{\rm^{13}CO}$  & $\tau_{\rm C^{18}O}$ & $N_{\rm ^{13}CO}$ & $N_{\rm C^{18}O}$ & $N_{\rm H_{2}}$($^{13}$CO) & $N_{\rm H_{2}}$(C$^{18}$O) & Name &$T_{\rm ex(CO)}$ & $\tau_{\rm^{13}CO}$  & $\tau_{\rm C^{18}O}$ & $N_{\rm ^{13}CO}$ & $N_{\rm C^{18}O}$ & $N_{\rm H_{2}}$($^{13}$CO) & $N_{\rm H_{2}}$(C$^{18}$O)\\ 
 & K & & & 10$^{15}$cm$^{-2}$ & 10$^{15}$cm$^{-2}$ & 10$^{22}$cm$^{-2}$ & 10$^{22}$cm$^{-2}$ & & K & & & 10$^{15}$cm$^{-2}$ & 10$^{15}$cm$^{-2}$ & 10$^{22}$cm$^{-2}$ & 10$^{22}$cm$^{-2}$  \\
\hline
G158.24-21.80  &  9.8(0.1)   &  0.6(0.01)  &  0.1(0.01) & 6.1(0.2)   &   0.6(0.2) &  0.6(0.2) &  0.4(0.1) &            G173.45-13.34  &  10.2(0.2)  &  1.0(0.01)  &  0.4(0.01) & 4.6(0.4)   &   1.0(0.1) &  0.4(0.3) &  0.7(0.1) \\
G158.37-20.72  &  16.2(0.4)  &  0.6(0.01)  &  0.1(0.01) & 14.0(0.9)  &   1.3(0.4) &  1.2(0.8) &  0.9(0.3) &            G173.89-17.64  &  10.7(0.4)  &  1.1(0.03)  &  0.3(0.01) & 6.9(0.9)   &   0.8(0.2) &  0.6(0.8) &  0.6(0.1) \\
G158.77-33.30  &  8.9(0.1)   &  0.8(0.05)  &  0.1(0.01) & 4.1(0.6)   &   0.5(0.2) &  0.4(0.6) &  0.4(0.1) &            G173.95-13.74  &  9.9(0.3)   &  2.7(0.11)  &  0.7(0.14) & 13.0(2.2)  &   2.7(0.8) &  1.2(2.0) &  1.9(0.6) \\
G158.86-21.60  &  14.0(0.4)  &  0.6(0.01)  &  0.1(0.01) & 9.5(0.9)   &   0.8(0.3) &  0.8(0.8) &  0.6(0.2) &            G174.06-15.81  &  13.2(0.2)  &  1.4(0.03)  &  0.3(0.02) & 17.0(0.9)  &   2.1(0.3) &  1.5(0.8) &  1.5(0.2) \\
G159.01-08.46  &  7.3(0.2)   &  1.5(0.05)  &  0.3(0.04) & 11.0(0.9)  &   1.0(0.3) &  1.0(0.8) &  0.7(0.2) &            G174.39-13.43  &  9.8(0.2)   &  1.2(0.02)  &  0.5(0.01) & 5.9(0.4)   &   2.3(0.2) &  0.5(0.3) &  1.6(0.1) \\
\hline
\end{tabular}
\end{sidewaystable}

\end{appendix}

\end{document}